%% file: example_paper.tex
\documentclass{article}

\usepackage{microtype}
\usepackage{graphicx}
\usepackage{subfigure}
\usepackage{booktabs} 

\usepackage{hyperref}
\usepackage{tikz}
\usepackage{pgfplots}
\usepackage{graphicx}
\usepackage{balance}
\usepackage{amssymb}
\usepackage[acronym]{glossaries}
\usepackage{xcolor}
\newacronym{VQ}{VQ}{vector quantizer}
\newacronym{SQ}{SQ}{scalar quantizer}
\newacronym{CL}{CL}{commitment loss}
\newacronym{STE}{STE}{straight-through estimator}
\newacronym{mSTE}{mSTE}{modified STE}
\newacronym{NA}{NA}{noise approximation}
\newacronym{MSE}{MSE}{mean-squared error}
\newacronym{MA-E}{MA-E}{mean-absolute embedding}
\newacronym{DAC}{DAC}{descript-audio-codec}
\usepgfplotslibrary{groupplots}
\usetikzlibrary{ positioning,shapes.geometric, arrows.meta, positioning, 3d, calc}
\usepackage{microtype}
\usepackage{graphicx}
\usepackage{subfigure}
\usepackage{booktabs} 
\usepackage{tikz}
\usetikzlibrary{shapes.geometric, arrows}
\usepackage{amsmath}
\usepackage{hyperref}

\usepackage[accepted]{icml2025}

\usepackage{amsmath}
\usepackage{amssymb}
\usepackage{mathtools}
\usepackage{amsthm}

\usepackage[capitalize,noabbrev]{cleveref}

\theoremstyle{plain}

\theoremstyle{definition}

\theoremstyle{remark}

\usepackage[textsize=tiny]{todonotes}

\icmltitlerunning{Efficient Evaluation of Quantization-Effects in Neural Codecs}

\begin{document}

\twocolumn[
\icmltitle{Efficient Evaluation of Quantization-Effects in Neural Codecs}




\begin{icmlauthorlist}
\icmlauthor{Wolfgang Mack}{yyy}
\icmlauthor{Ahmed Mustafa}{yyy}
\icmlauthor{Rafał Łaganowski}{yyy}
\icmlauthor{Samer Hijazy}{yyy}
\end{icmlauthorlist}

\icmlcorrespondingauthor{Wolfgang Mack}{womack@cisco.com}
\icmlaffiliation{yyy}{Cisco Systems, Inc., \{womack, ahmmusta, rlaganow, hijazy\}@cisco.com}

\vskip 0.3in
]



\printAffiliationsAndNotice{}  
\begin{abstract}

Neural codecs, comprising an encoder, quantizer, and decoder, enable signal transmission at exceptionally low bitrates. Training these systems requires techniques like the straight-through estimator, soft-to-hard annealing, or statistical quantizer emulation to allow a non-zero gradient across the quantizer. Evaluating the effect of quantization in neural codecs, like the influence of gradient passing techniques on the whole system, is often costly and time-consuming due to training demands and the lack of affordable and reliable metrics. This paper proposes an efficient evaluation framework for neural codecs using simulated data with a defined number of bits and low-complexity neural encoders/decoders to emulate the non-linear behavior in larger networks. Our system is highly efficient in terms of training time and computational and hardware requirements, allowing us to uncover distinct behaviors in neural codecs. We propose a modification to stabilize training with the straight-through estimator based on our findings. We validate our findings against an internal neural audio codec and against the state-of-the-art descript-audio-codec.

\end{abstract}

\section{Introduction}
\label{submission}
Codec systems typically consist of an encoder, a quantizer, and a decoder. The encoder transforms high-dimensional data such that it can be quantized, i.e. represented with a typically much smaller number of discrete values. Storing or transmitting data in the quantized representation is crucial in modern multi-media environments as it significantly reduces storage and transmission requirements and typically does not reduce perceived quality like audio clarity or video sharpness significantly (e.g., \cite{4604423,muller24c_interspeech}).  Subsequently, the decoder reconstructs the data from the quantized representations. 

 Classical coding methods, such as JPEG \cite{jpegstandard}, MP3 \cite{mp3}, H.264 \cite{h264}, and  Opus \cite{opus} are based on signal processing and perceptual models  \cite{SpeechCodingBase,bhaskaran1995image}. Despite their historical success, these techniques struggle to achieve high efficiency at very low bitrates, where signal fidelity becomes a critical challenge (e.g., \cite{Zeghidour2022}).

With the rise of deep learning, classical methods have been combined with neural networks. Neural vocoders, for example, have been employed to reconstruct high-fidelity signals from bitstreams of classical methods, achieving unprecedented performance in speech and audio compression \cite{Kleijn2018, Klejsa2019, Garbacce2019,  Valin2019a,9632750}. Neural networks are also used as post-processing tools to mitigate compression artifacts, thereby enhancing the perceptual quality of the decoded signals \cite{Zhao2019, Korse2020, Biswas2020, Korse2022, Buthe2024}. 

End-to-end neural compression systems that bypass traditional coding elements have emerged as an alternative. These systems learn to encode, quantize, and decode signals directly from data and can be trained end-to-end \cite{ VanDenOord2017,  balle2016endtoend,Agustsson2019,Zeghidour2022, Defossez2022, NSVQ,Vali2023,  Kumar2023, Ai2024, Mentzer2024, Brendel2024}. However, training such systems requires tricks like the straight-through estimator \cite{VanDenOord2017,bengio2013ste}, soft-to-hard annealing \cite{agustsson2017soft,Jang2017,maddison2017gumbel}, or statistical quantizer emulation \cite{balle2016endtoend,9242247,NSVQ,Vali2023,Brendel2024} to allow a non-zero gradient to pass over the quantizer.  

Evaluating neural codecs is nontrivial, costly, and time-consuming because of the required training and is often not practical because of the lack of cheap and reliable metrics. Evaluation points for neural codecs, such as quantization error or decoder output, face challenges due to the non-linear systems involved. For example, a small quantization error can lead to a significant reconstruction error and vice versa. Metrics for assessing decoder outputs in the audio, image, or video domains are often either subjective and resource-intensive or objective but unreliable. For example, MUSHRA (Multiple Stimuli with Hidden Reference and Anchor) is a subjective evaluation method widely regarded as the gold standard for audio quality assessment. Although highly effective in capturing perceptual quality, MUSHRA tests require human participants, limiting their scalability to a small number of files and increasing the associated costs. Furthermore, MUSHRA scores often show limited correlation with objective metrics, a disparity that becomes particularly pronounced when evaluating generative models such as neural codecs.

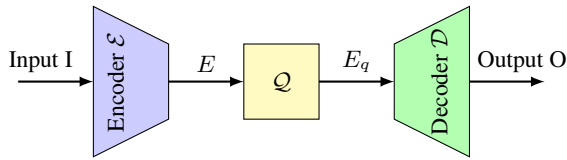
\begin{figure}[t]
    \centering
    \begin{tikzpicture}[>=latex, font=\small]

        \draw[fill=blue!20] (-4, 1) -- (-3, 0.5) -- (-3, -0.5) -- (-4, -1) -- cycle;
        \node[rotate=90] at (-3.7, 0) {Encoder $\mathcal{E}$};

        \draw[fill=yellow!30] (-2, 0.5) rectangle (-1, -0.5);
        \node at (-1.5, 0) {$\mathcal{Q}$};

        \draw[fill=green!30] (0, 0.5) -- (1, 1) -- (1, -1) -- (0, -0.5) -- cycle;
        \node[rotate=90] at (0.6, 0) {Decoder $\mathcal{D}$};

        \node[above] at (-4.7, 0) {Input I};
        \node[above] at (1.7, 0) {Output O};

        \node[above] at (-2.5, 0) {$E$};         
        \node[above] at (-0.5, 0) {$E_q$}; 

        \draw[->, thick] (-5, 0) -- (-4, 0);    
        \draw[->, thick] (-3, 0) -- (-2, 0);    
        \draw[->, thick] (-1, 0) -- (0, 0);     
        \draw[->, thick] (1, 0) -- (2, 0);      

    \end{tikzpicture}
    \caption{A neural codec system with an encoder $(\mathcal{E})$ mapping the input to embeddings $E$, a quantizer ($\mathcal{Q}$) mapping $E$ to the quantized version $E_q$. The decoder $(\mathcal{D})$ maps $E_q$ to the output. }
    \label{fig:neural_codec}
\end{figure}
In this paper, we propose an evaluation framework that enables efficient investigation and evaluation of quantizers in neural codecs in a controlled way. The proposed framework consists of a simple input and target data simulation method to train a neural codec. The input data is designed to have a specified number of bits. Designing the data in that way enables the identification of the minimum number of bits required by the quantizer to reconstruct the target perfectly. The target is a rotated input version containing the same number of bits. The input and target data design are based on quantized noise processes, which ensures a resource-efficient, low-cost simulation framework. In addition to the data simulation, we propose to use a low-complexity neural codec. That way, we emulate the highly non-linear behavior before and after the quantizer in a large network by simultaneously keeping hardware requirements and training cost/time extremely low. For training the neural codec, we propose to use an interpretable cost function like the \gls{MSE} between target and estimate to evaluate the quantizer performance. The proposed loss is in contrast to real codecs, where the estimate is hard to assess regarding quality. Using the proposed framework, we evaluate fundamental properties of neural codecs by comparing training using statistical quantizer emulation to training using the \gls{STE} with and without  \gls{CL}. We find similarities between both approaches and propose a modification to stabilize training with the \gls{STE}. Finally, we verify our findings by repeating selected experiments with an internal audio codec and \gls{DAC} \cite{Kumar2023}.   

The remainder of the paper is structured as follows. Fundamental mathematical notations are introduced in Section~\ref{sec:fundamentals}. The proposed method is presented in Section~\ref{sec:prop}, followed by the experimental parameters in Section~\ref{sec:paras}. In Section~\ref{sec:Peval}, experiments are covered. Finally, Section~\ref{sec:conclusion} contains a brief conclusion.
\section{Fundamentals}
\label{sec:fundamentals}

We consider an encoder-quantizer-decoder system, as illustrated in Figure~\ref{fig:neural_codec}. The encoder $\mathcal{E}$ maps the input $I$ to embeddings $E \in \mathbb{R}^{F \times N}$, where $F$ represents the feature dimension and $N$ the frame dimension. A quantization module $\mathcal{Q}$ then maps $E$ to its quantized form $E_q$. Finally, the decoder $\mathcal{D}$ processes $E_q$ to estimate the output $O$. Depending on the application, $I$ and $O$ could be images, audio, or other signals. Without loss of generality, we refrain from specifying the exact nature of these signals. Instead, we focus on analyzing the quantization module $\mathcal{Q}$ and its effects on the neural codec. In particular, we examine its influence on the relationship between $E$ and $E_q$, as well as the gradient flow through $\mathcal{Q}$\footnote{Note that $\mathcal{Q}$ is embedded in the highly non-linear encoder-decoder system. Consequently, analyzing quantizers in the form of their quantization error is insufficient to evaluate their performance in neural networks, as a small quantization error might lead to a large output error and vice versa. }.

The gradient flow over $\mathcal{Q}$ poses a fundamental challenge: in most quantizer designs, $\mathcal{Q}$ is non-differentiable because it maps nearly continuous embeddings $E$ to discrete values or vectors in $E_q$. This non-differentiability prevents loss functions defined after $\mathcal{Q}$ from directly updating the encoder weights. Various techniques have been proposed to overcome this limitation,  including (1) the \gls{STE} \cite{VanDenOord2017,bengio2013ste}, (2) statistical training using noise-based approximations for the quantization error \cite{balle2016endtoend,9242247,NSVQ,Vali2023,Brendel2024}, and (3) soft quantization \cite{agustsson2017soft,Jang2017,maddison2017gumbel}.
\begin{figure}[t]
    \centering
    \newsavebox\myboxa
    \savebox\myboxa{%
    \begin{tikzpicture}
    \begin{axis}[scale=0.3,
        hide axis,
        colormap/cool,title style={yshift=-0.8cm}, zmin=0, zmax=1]
        \addplot3[surf,domain=-3:3,domain y=-3:3,] 
            {exp(-( (x)^2 + (y)^2)/3 )};
    \end{axis}
    \end{tikzpicture}%
    }

    \begin{tikzpicture}
    \tikzstyle{branch}=[fill,shape=circle,minimum size=3pt,inner sep=0pt]
    \node[label={[label distance=-0.5cm]above:{$I \in \mathbb{R}^{P\times P}$}}] (input) {\usebox\myboxa};
    
    \node[draw, rectangle, right=of input] (Sample) {\rotatebox{90}{Sampling}};
    
    \node[draw, rectangle,  right=of Sample] (Q) {\rotatebox{90}{Scalar Quantization}};
    
    \node[right=of Q, circle, draw] (o1) {$\odot$}; 
    \node[above= of o1] (rot) {Rotation Matrix $Q$};
    \node[right=of o1] (o2) {}; 

    \draw[->] (input) -- (Sample) ;  
    \draw[->] (Sample) -- (Q) node[above,midway]{$X$};  
    \draw[->] (Q) -- (o1) node[above,midway]{$X_q$}; 
    \draw[->] (o1) -- (o2)node[above,midway]{$Y$};;  
    \draw[->] (rot) -- (o1); 
    \end{tikzpicture}
       \caption{Proposed data generation pipeline. A Gaussian noise process with a $P\times P$ identity covariance matrix is sampled to obtain $X$. Each element in $X$ is quantized via scalar quantization to obtain the network target $X_q$. A rotation matrix is applied to $X_q$ to obtain the network input $Y$.  }
    \label{fig:datagen}
\end{figure}
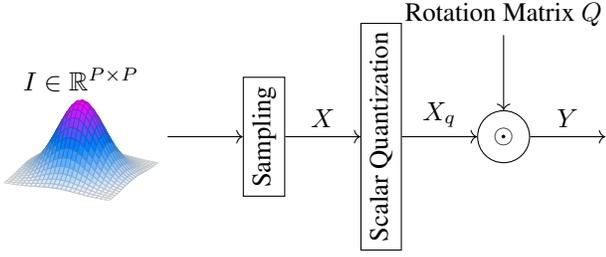

In the case of the \gls{STE}, the decoder input $\mathcal{D}^{\text{STE}}_{\text{in}}$ during training is defined as
\begin{equation}
\mathcal{D}^{\text{STE}}_{\text{in}} = E + \text{sg}[\underbrace{E_q - E}_{Q_e}],
\label{equ:STE}
\end{equation}
where $Q_e$ is the quantization error, and $\text{sg}[\bullet]$ stops gradient tracking for $\bullet$ \footnote{Note that  $\text{sg}[(E_q-E)]$ is crucial, without it the decoder input would be $E_q = E +(E_q-E)$ which has gradients of $0$ w.r.t. $E$.}. During the forward pass, $\mathcal{D}^{\text{STE}}_{\text{in}}$ equals $E_q$, while in the backward pass, the gradient through $\mathcal{Q}$ simplifies to
\begin{equation}
\frac{\partial \mathcal{D}^{\text{STE}}_{\text{in}}}{\partial E} = 1^{F\times N}
\label{equ:derSTE}
\end{equation}
due to the sg in (\ref{equ:STE}).  According to \cite{VanDenOord2017}, a \gls{CL} is required for training to enforce $E_q$ and $E$ to be close, i.e., 
\begin{equation}
    \mathcal{L}_\text{CL} = \sum_{F,N}\frac{(E-\text{sg}[E_q])^2}{F\cdot N}.
\end{equation}
In (\ref{equ:STE}), the term $Q_e$ acts as an additive noise component not connected to the computational graph. Training using \gls{NA} follows a similar philosophy as training using \gls{STE} in the sense that a noise component is added. In \gls{NA}, noise $U $ is added to $E$ to simulate the quantization process, i.e., 
\begin{equation}
   \mathcal{D}^\text{NA}_\text{in} = E + U. 
   \label{equ:classicNA}
\end{equation}
Note that in contrast to using the \gls{STE}, the distribution of $U$ is a design choice. For example, it can be set to produce a specific embedding-to-noise ratio. The derivative of $\mathcal{D}_\text{in}^{\text{NA}}$ w.r.t. $E$ can be written as
\begin{equation}
    \frac{\partial \mathcal{D}^{\text{NA}}_{\text{in}}}{\partial E} = 1^{F\times N} + \frac{\partial U}{\partial E}.
    \label{equ:derUNA}
\end{equation}
After training using \gls{NA}, a quantizer replaces the noise addition for inference. Typical quantization modules used in neural codecs are based on a  \gls{SQ} (e.g., \cite{Brendel2024,Mentzer2024}) or a \gls{VQ} (e.g., \cite{Zeghidour2022, NSVQ}).

Soft quantization (3) offers a differentiable alternative by replacing the hard, discrete mapping of $\mathcal{Q}$ with a continuous approximation during training. The quantization of $E$ is modeled as a weighted average of discrete codebook entries or quantization levels, allowing gradient propagation through $\mathcal{Q}$. During training, the weights are forced to increasingly approximate a one-hot vector, enabling a gradual annealing from continuous to discrete values. The soft assignment probabilities are replaced during inference by a hard one-hot assignment.
\section{Proposed Method}
\label{sec:prop}
Comparing the effect of quantizers and their emulations in a neural codec is not trivial due to the lack of reliable metrics and costly due to long and computationally heavy trainings. We propose a low-complexity method to evaluate the impact of quantization on encoder-quantizer-decoder neural networks. In particular, we propose the use of surrogate data and a low-complexity surrogate model to investigate the effects of quantization. 
\subsection{Data Simulation}
\label{subsec:surrdata}

In this study, we propose using simulated input and target data for neural codec training, which are both (1) easy to generate and cost-effective and (2) contain a fixed amount of information measured in bits. This approach allows us to determine the minimum bit requirement of $\mathcal{Q}$ for perfect reconstruction while eliminating complex loss functions, thus enabling clearer comparisons between training sessions. By selecting simplified data, we also reduce the impact of encoder/decoder capacity on observed losses. A scheme of the data simulation is depicted in Figure~\ref{fig:datagen}.

We define $X \in \mathbb{R}^{P \times N}$, which consists of $N$ samples from $P$ Gaussian noise processes. To achieve objective (1), we quantize $X$ to obtain $X_q$, where the number of bits used in quantization determines the information of $X_q$. Recognizing that real-world signals, such as audio, exhibit sample correlations, we propose correlating the processes $P$ with each other via

\begin{equation}
    Y = Q \odot X_q,
\end{equation}

where $Y \in \mathbb{R}^{P \times N}$ serves as the input data for the neural codec, and $Q \in \mathbb{R}^{P \times P}$ is a rotation matrix, with $\odot$ denoting the matrix product. Note that $Q$ is derived from the QR decomposition of a $P \times P$ matrix containing samples of a white Gaussian noise process. Consequently, $Q$ only applies a rotation and does not change the eigenvalues of $X_q$, ensuring that $Q^{-1} = Q^T$, which is easily invertible.

We define the target of the neural codec as $X_q$. The proposed data definition satisfies requirements (1) and (2), as both $X_q$ and $Y$ carry a specific amount of information, are easy to simulate, and the mapping of $Y$ to $X_q$ is straightforward with

\begin{equation}
    X_q = Q^T \odot Y
\end{equation}
The simple target mitigates the effect of encoder/decoder capacity on the losses as the neural codec task is to get the estimate $\widehat{X}_q$ of $Y$, i.e., 
\begin{equation}
    \widehat{X}_q = \mathcal{D}(\mathcal{Q}(\mathcal{E}(Y))).
\end{equation}

\subsection{Neural Codec and Training}
\label{subsec:surrmodel}
The codec used for evaluation must meet the following criteria: (1) it should be fast to train and maintain low complexity, and (2) it should incorporate non-linearities similar to those in large networks such that the quantizer operates in a highly non-linear system. This is crucial for codec evaluation, as in a non-linear system, a small/large quantization error does not have to correspond to a small/large output error. 

We propose constructing the encoder-decoder using fully connected layers with skip connections to achieve these objectives. This architecture enables the model to effectively capture complex relationships within the data. The model operates on a frame basis, mapping $Y[:,n]$ to $X_q[:,n]$. Training is conducted using the \gls{MSE} loss
\begin{equation}
    \mathcal{L}_\text{MSE} = \frac{1}{F \cdot N}\sum_{F,N} (X_q - \widehat{X}_q)^2.
    \label{equ:MSELoss}
\end{equation}

\subsection{Modified Straight-Through Estimator}
We propose a \gls{mSTE} that stabilizes training neural codecs even when no \gls{CL} is used. As stated in \cite{VanDenOord2017}, the \gls{STE} leads to an unstable system as $E$ is constantly growing when no \gls{CL} is used during training. To stabilize the \gls{STE}, we propose to multiply the quantization noise with a modifier, i.e.,
\begin{equation}
    \mathcal{D}_\text{in}^{\text{mSTE}} = \underbrace{E + \text{sg}[Q_e]}_{\text{STE}}\cdot \underbrace{\frac{\sigma_{Q_e}}{\text{sg}[\sigma_{Q_e}]}}_{\text{modifier}},
    \label{equ:mSTE}
\end{equation}
where $\sigma_{Q_e} \in \mathbb{R}_{\geq 0}$ is the standard deviation of $Q_e=E_q-E$. The multiplication with the modifier is similar to the reparametrization-trick in \cite{Kingma2014}, where a noise process is connected via multiplication with an estimated standard deviation to a neural network graph. In (\ref{equ:mSTE}), we connect the quantization noise $Q_e$ to the graph by multiplying with $\sigma_{Q_e}$. To avoid changing the decoder input in the forward pass, we divide by $\text{sg}[\sigma_{Q_e}]$. That way, the modifier term is $1$ in the forward pass. Consequently, the decoder input is $E_q$ in the forward pass.  In contrast, the modifier changes the backward pass such that 
\begin{equation}
\frac{\partial \mathcal{D}^{\text{mSTE}}_{\text{in}}}{\partial E} = \mathbf{1}^{F \times N} + 
\text{sg} \left[ \frac{Q_e}{\sigma_{Q_e}} \right] \cdot\frac{\partial \sigma_{Q_e}}{\partial E} .
\label{equ:derMSTE}
\end{equation}
As $\sigma_{Q_e}$ depends on $E$ and is part of the computational graph, the proposed modification changes the update of the encoder. With the proposed modifier, $Q_e$ is normalized in the backward pass with $\sigma_{Q_e}$ in (\ref{equ:derMSTE}), stabilizing the gradients. We hypothesize that the growth of $E$ when using \gls{STE} without \gls{CL} reported in \cite{VanDenOord2017} is caused by this missing connection of $Q_e$ to the computational graph during backpropagation. Simply, the model tries to maximize the embedding-to-noise ratio by increasing the norm of $E$. Clearly, the model cannot succeed because a larger $E$ typically leads to a larger $Q_e$, which leads to an unstable process. The \gls{CL} stops the growth of $E$ at the cost of having a loss with a trivial solution when $E=0^{F\times N}$ (i.e., the information in $E$ is zero). Using the proposed \gls{mSTE}, we assume that \gls{CL} is no longer required for training.
\section{Experimental Parameters}
\label{sec:paras}
Here we present the experimental parameters we set and the data we used.

\textbf{Data}:
For $X$, we set $P=30$ and $N=2000$. We sample $X$ of a white Gaussian noise process with zero mean and variance one. For $X_q$ we use $2$~bits per value ($60$~bits per frame) and \gls{SQ} with fixed levels $\{-1.5,-0.5,0.5,1.5\}$. 

\textbf{Neural Network Plus Training} Encoder and decoder consist of $3$ fully connected layers each. We use skip connections except in the last encoder layer and the first and last decoder layer. The input and output dimensions of each layer are set to $30$. We use no activation at the encoder output and a PReLU activation for all other layers. We train for $100$ epochs, where one epoch consists of $2000$ updates with one $E$ of shape $30\times 2000$ each. We use Adam with a learning rate of $1e-4$ as an optimizer. 

\textbf{Quantizers/Quantization Emulation}: In the evaluation, we consider two training techniques, (1) \gls{SQ} and (2) NA. In both cases, we use a latent dimension of $F=30$. 

For \gls{SQ}, we use $2$~bits per value ($60$~bits per frame) and \gls{SQ} with fixed levels $\{-1.5,-0.5,0.5,1.5\}$. We mark models trained using \gls{SQ} as $\text{SQ}^\bullet_\star$ where $\star \in \{ \text{STE}, \text{mSTE}\}$ marks the gradient estimator and $\bullet \in \{ ,\text{CL}\}$ specifies whether \gls{CL} with weight of $0.1$ was used for training.

For NA, we construct $U$ by scaling a white Gaussian noise process $\mathcal{N}$ with zero mean and variance one with $\alpha \in \mathbb{R}$, i.e., \begin{equation}
U = \alpha \cdot \sigma_E \cdot \mathcal{N}(0,1) 
\label{equ:alphauat}
\end{equation}
with $20\cdot\log_{10}(\alpha\cdot \sigma_E)\in [0,8]$~dB controlling the embedding-to-noise ratio and $\sigma_E \in \mathbb{R}_{\geq 0}$ is the standard deviation of $E$. We consider two cases for NA, (1) NA as in (\ref{equ:classicNA}) and (2) $\text{NA}_\text{det}$ with detached noise, i.e., 
\begin{equation}   \mathcal{D}_\text{in}^{\text{NA}_\text{det}} = E + \text{sg}[U] 
      \label{equ:NAdet}
\end{equation}
where $\mathcal{D}^\text{$\text{NA}_\text{det}$}_\text{in}$  is the decoder input for $\text{NA}_\text{det}$. In $\text{NA}_\text{det}$, the gradients of $U$ w.r.t. $E$ are zero. For NA, the gradients of $U$ w.rt. $E$ depend on $\sigma_E$.

\textbf{Descript-Audio-Codec (DAC)}
We repeat selected experiments using the \gls{DAC} \cite{Kumar2023} implementation from \cite{descript_audio_codec}. For training, we use the dev-clean subset of the LibriTTS dataset \cite{zen19_interspeech} consisting of $8.97$ hours ($20$ male and $20$ female English speakers) and train the model at 16kHz sampling rate using the default parameters (without adversarial loss), beside activating/deactivating \gls{CL} and introducing \gls{mSTE} in the code. We refer to the respective models as $\text{DAC}^\text{CL}_\text{STE}$, $\text{DAC}^\text{CL}_\text{mSTE}$,  $\text{DAC}_\text{STE}$, and $\text{DAC}_\text{mSTE}$.

\textbf{Metrics} For evaluation purposes, we present over the training the \gls{MSE} in (\ref{equ:MSELoss}) and the mean-absolute of $E$, i.e., 
\begin{equation}
\text{MA-E} = \frac{ \lVert E\rVert_1}{F\cdot N},
\label{equ:MAE}
\end{equation}
where $\lVert E\rVert_1$ is the L1-norm of $E$.

\section{Evaluation of Selected Properties of Neural Codec Systems}
\label{sec:Peval}
\label{sec:eval}
\begin{table}[t!]
    \centering
    \caption{Memory and training time for different trainings with the proposed framework compared to DAC \cite{Kumar2023} (trained without adversarial loss) on a single P40 GPU.}
    \begin{tabular}{lcc}
        \toprule
        & Proposed Framework  & DAC \\
        \midrule
        Mem. & $< 400$ MB & 15.5 GB \\
        Time & $< 1$h  & $\approx$ 1 Week \\
        \bottomrule
    \end{tabular}
    \label{tab:mem_time}
\end{table}
\begin{figure}[t]
    \centering
    \input{Figures/Plot0loss}
    \caption{Training \gls{MSE} and \gls{MA-E} when using a no quantizer and the standard training using the \gls{STE} with \gls{CL}.}
    \label{fig:enter-label}
\end{figure}
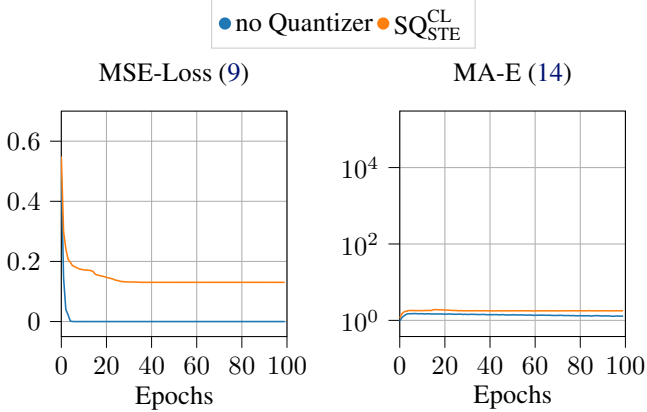
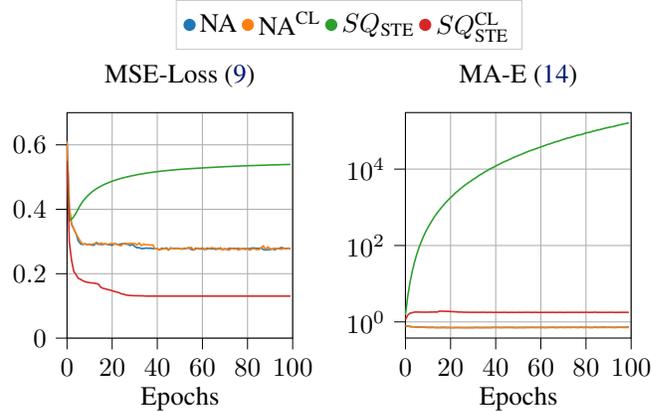
\begin{figure}[t]
    \centering
    \input{Figures/Plot1loss}

    \caption{Training \gls{MSE} and \gls{MA-E} when using \gls{NA} or the \gls{STE} for training with and without \gls{CL}. Note that the \gls{MA-E} curves of NA and $\text{NA}^\text{CL}$ are overlapping.}
    \label{fig:NIvsSTE}
\end{figure}

In evaluating the proposed framework, we focus on fundamental properties of neural codecs. We note that further investigations not part of the article can easily and quickly be accomplished with the proposed method. From our point of view, these are but are not limited to: (1) Network layers, activations, or regularizers directly before or after the quantizer (e.g., dropout leads to correlated features such that a \gls{VQ} might work better than a \gls{SQ}), (2) Other gradient estimators like ReinMax \cite{liu2024bridging}, or SPIGOT \cite{peng2018backpropagating}, (3) different quantizers and bitrates including soft quantizers \cite{agustsson2017soft,Jang2017,maddison2017gumbel}.

A comparison of training time and required GPU memory is given in Table~\ref{tab:mem_time}. The proposed system takes less than an hour to train and requires less than $400$~MB of GPU memory. Consequently, multiple systems can be tested in parallel, enabling rapid prototyping. Compared to full-scale neural codecs like \gls{DAC}, temporal and hardware requirements are significantly reduced, leading to a significantly reduced cost.

\subsection{Experiments with the Proposed Evaluation Framework}

\textbf{No Quantizer vs. STE}: First, in Figure~\ref{fig:enter-label}, we compare the training of the proposed system without a quantizer and with $\text{SQ}_\text{STE}^\text{CL}$. As expected, the loss without quantizer approaches zero very quickly. For $\text{SQ}_\text{STE}^\text{CL}$, the model is able to reduce the loss; however, it converges to an \gls{MSE} of $\approx 0.13$. Theoretically, when $\mathcal{Q}$ has the same number of bits or more than the input, perfect reconstruction is possible. For $\text{SQ}_\text{STE}^\text{CL}$, both an input frame and the bits of the quantizer are the same (60~Bits). When doubling the bits in the quantizer to 120, the \gls{MSE} loss yields comparable results to a quantizer-free system.  The \gls{MA-E} in Figure~\ref{fig:enter-label} is stable for all models, as it converges to a fixed value over the epochs. A stable, non-diverging system is important. Otherwise, the input of the decoder constantly changes, and the model cannot learn.

\textbf{no CL vs. CL}: In Figure~\ref{fig:NIvsSTE} we compare \gls{NA} vs. \gls{STE} with and without \gls{CL}. The \gls{MSE} and \gls{MA-E} values of NA and $\text{NA}^\text{CL}$ are comparable over training, showing that \gls{CL} has no effect on them. For $\text{SQ}_\text{STE}^\text{CL}$ and $\text{SQ}_\text{STE}$ the \gls{MSE} and \gls{MA-E} show a strong difference. Without \gls{CL}, the encoder output $E$ grows unboundedly over training, whereas when using \gls{CL}, the \gls{MA-E} converges. In \cite{VanDenOord2017}, the authors mention the same behavior for training a neural codec using the \gls{STE} without \gls{CL}. The growing $E$ translates to a diverging loss for $\text{SQ}_\text{STE}$. The different effect of  \gls{CL} on \gls{NA} and \gls{STE} is remarkable. In the following, we analyze what incentivises the encoder to produce larger and larger $E$ for \gls{STE} without \gls{CL} but not for \gls{NA}.

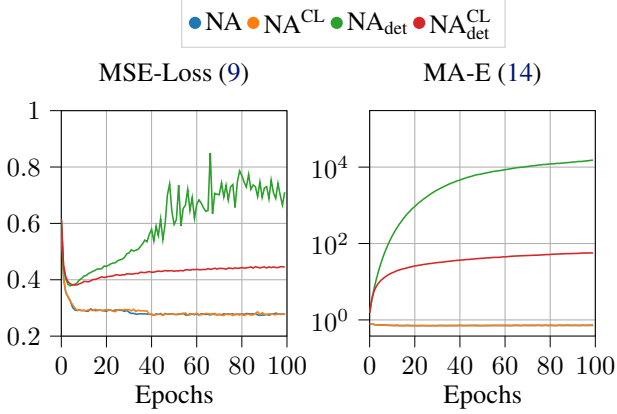
\begin{figure}[t]
    \centering
    \input{Figures/Plot2loss}
    \caption{Training \gls{MSE} and \gls{MA-E} when using \gls{NA} for training with and without \gls{CL}. We compare training using detached and attached noise as in (\ref{equ:NAdet}) and (\ref{equ:classicNA}), respectively. In the \gls{MA-E}, NA and $\text{NA}_\text{det}$ overlap.}
    \label{fig:NIdetvsat}
\end{figure}
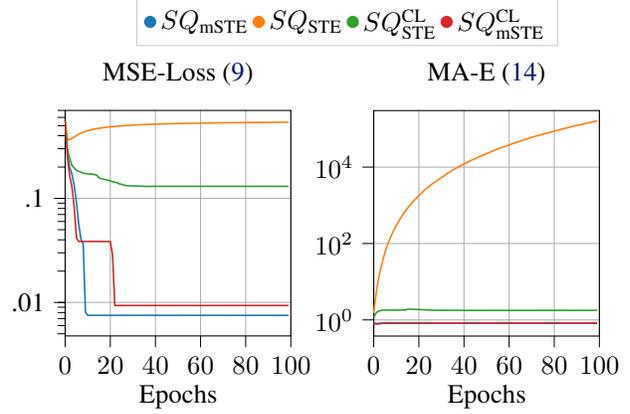
\begin{figure}[t]
    \centering
    \input{Figures/Plot3loss}
    \caption{Training \gls{MSE} and \gls{MA-E} when using the \gls{STE} with and without \gls{CL}. \gls{mSTE} is the proposed modification of the \gls{STE} from (\ref{equ:mSTE}). Note that the blue and the red curve overlap in the \gls{MA-E}.}
    \label{fig:stevsmodste}
\end{figure}
\textbf{Detached vs. Attached Noise Approximation}: Investigating the growth of $E$ further for the $\text{STE}$, we analyze the effect of attached and detached noise $U$ for NA. As mentioned in Section~\ref{sec:paras}, the noise level $U$ depends on a fixed embedding-to-noise ratio. Inserting (\ref{equ:alphauat}) in (\ref{equ:derUNA}), we obtain 
\begin{equation}
    \frac{\partial \mathcal{D}^{\text{NA}}_{\text{in}}}{\partial E} = 1^{F\times N} + \alpha \cdot \mathcal{N}(\mu=0, \sigma=1)\cdot\frac{\partial \sigma_E }{\partial E},
    \label{equ:derUNA2}
\end{equation}
effectively connecting the noise level to the standard deviation of $E$ in the encoder update. In contrast, when using \gls{STE} or detached noise, 
\begin{equation}
    \frac{\partial \mathcal{D}_\text{in}^{\text{NA}_\text{det}}}{\partial E} =    \frac{\partial \mathcal{D}_\text{in}^{\text{STE}}}{\partial E} = 1^{F\times N},
\end{equation}
which does not connect the the noise ($Q_e$ or $U$) to the computational graph. Effectively, for \gls{STE} and $\text{NA}_\text{det}$, the decoder observes noisy encoder outputs $E_q$ where the noise (either $U$ or $Q_e$) is not part of the computational graph. Consequently, the encoder weights are changed so that the embedding-to-noise ratio is maximized, leading to a divergence of $E$. Figure~\ref{fig:NIdetvsat} depicts the \gls{MSE} and the \gls{MA-E} over training for \gls{NA}. Clearly, we see the expected growth of $E$ when training with $\text{NA}_\text{det}$.  Also for $\text{NA}_\text{det}^{\text{CL}}$ and $\text{NA}_\text{det}$ we see that when the noise is detached, the \gls{CL} is required to stabilize the training. Interestingly, using attached noise without \gls{CL} seems to be advantageous over detached noise with \gls{CL}, as the \gls{MSE} of $\text{NA}_\text{det}^{\text{CL}}$ is higher than of NA. Looking at the \gls{MA-E}, the norms of $\text{NA}_\text{det}^{\text{CL}}$ keep increasing, hinting at a non-sufficient weighting of \gls{CL} (used weighting is $0.1$). 

\textbf{Proposed mSTE vs. STE}: The gradients of the proposed \gls{mSTE} in (\ref{equ:derMSTE}) and of training using $\text{NA}$ in (\ref{equ:derUNA2}) exhibit explicit similarities. In particular, noise ($Q_e$ or $U$) is connected to the computational graph through a standard deviation that is based on the encoder layers ($\sigma_{Q_e}$ or $\sigma_E$). We hypothesize that this connection of the noise to the computational graph hinders the model from maximizing the embedding-to-noise ratio by increasing $E$ as this would simultaneously lead to a growth of the noise. Figure~\ref{fig:stevsmodste} depicts a comparison of the \gls{STE} and \gls{mSTE} with and without \gls{CL}. As expected, training using \gls{mSTE} does not require a \gls{CL} to have stable norms of $E$. Moreover, the \gls{MA-E} of $\text{SQ}^\text{CL}_\text{mSTE}$,  $\text{SQ}_\text{mSTE}$ and $\text{SQ}^\text{CL}_\text{STE}$ are stable, whereas the \gls{MA-E} of $\text{SQ}_\text{STE}$ diverges.  For \gls{mSTE}, the \gls{MSE} is very low compared to \gls{STE}. Interestingly, for $\text{SQ}^\text{CL}_\text{mSTE}$, we see a step function of the loss and a slower convergence compared to $\text{SQ}_\text{mSTE}$. We assume that the \gls{CL} makes the fine adjustment of $E$ difficult for low bitrates as they are drawn towards the respective quantization level. 

\subsection{Experiments with Neural Audio Codecs}
    \begin{figure}[t!]
    \centering
    \input{Figures/PlotXcodecf_norms}
    \caption{Training \gls{MA-E} over epochs for an internal neural audio codec (trained on audio). The dashed lines are the final values of trained models. We show the evolvement over the epochs for $\text{NA}_\text{det}$. After $32$ epochs, the model training crashed as $E$ grew too large. The other two models were trained for more than $1000$ epochs.}
    \label{fig:xcodec}
\end{figure}
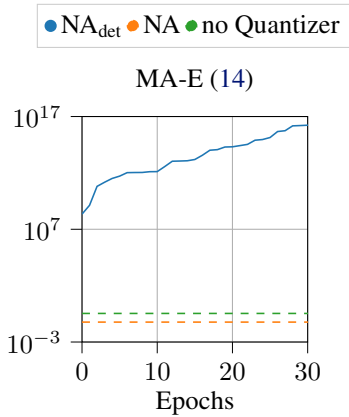
We repeat selected experiments with an internal audio codec and with \gls{DAC} trained using audio data to show the consistency of our findings.

\textbf{Effect of NA on Internal Speech Codec}: First, we evaluate the effect of \gls{NA} when training an internal model for speech coding. The embedding norms are depicted in Figure~\ref{fig:xcodec}. When no quantizer or NA is used, the final \gls{MA-E} are in a similar range to the \gls{MA-E} of the low-complexity model in Figure~\ref{fig:enter-label} and Figure~\ref{fig:NIdetvsat}, respectively. When detaching the added noise in $\text{NA}_\text{det}$, $E$ keeps growing till the training crashes after approximately 30~epochs. This result is consistent with the growth of $E$ for the low complexity model in Figure~\ref{fig:NIdetvsat}. 

\textbf{Effect of STE/mSTE on Descript-Audio-Codec (DAC)}:
 \begin{figure}[t!]
    \centering
    \input{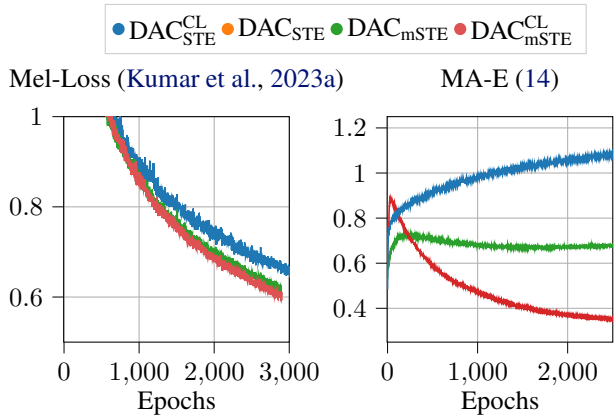}
    \caption{Mel-loss over training for \gls{DAC}. Each value is the average of $50$ updates. For $\text{DAC}_\text{STE}$, the curves are out of the range of the plot (Mel-Loss$>12$, \gls{MA-E}$>1e15$). }
    \label{fig:dacmel}
\end{figure}
We trained \gls{DAC} with and without \gls{CL} using \gls{STE} and \gls{mSTE}. The results for the Mel-Loss are illustrated in Figure~\ref{fig:dacmel}. Using $\text{DAC}_\text{STE}$, the loss rapidly diverges with an \gls{MA-E} $>1e15$ from the beginning, and the training crashes. As expected, $\text{DAC}_\text{mSTE}$ does not exhibit a diverging loss. Comparing the trainings to their counterparts with \gls{CL}, $\text{DAC}^\text{CL}_\text{STE}$ and $\text{DAC}^\text{CL}_\text{mSTE}$, both losses converge. The losses of $\text{DAC}^\text{CL}_\text{mSTE}$ and $\text{DAC}_\text{mSTE}$ are very close over the whole training. The similarity of the loss curves matches the results of the low-complexity model visualized in Figure~\ref{fig:stevsmodste}. Interestingly, the longer the network is trained, the larger the improvement of \gls{mSTE} over \gls{STE}. We assume that the stability of the \gls{MA-E} when using \gls{mSTE} contributes to that improvement. The \gls{MA-E} in Figure~\ref{fig:dacmel} show that, although marginal, $E$ is still growing for $\text{DAC}^\text{CL}_\text{STE}$ over the epochs. For $\text{DAC}^\text{CL}_\text{mSTE}$, the \gls{MA-E} is being reduced over the epochs. This can be explained as the \gls{CL} can simply be optimized by scaling down $E$ and $E_q$ as it is \gls{MSE}-based. For $\text{DAC}_\text{mSTE}$, \gls{MA-E} reaches a stable state very early in training ($\approx 100$~Epochs). Overall, the proposed \gls{mSTE} outperforms the \gls{STE} when using \gls{DAC} or the proposed efficient evaluation framework.
\section{Conclusion}
\label{sec:conclusion}
We proposed an efficient analysis framework for neural codecs based on simulated data and small neural networks. Training the proposed system required less than one hour on a GPU compared to several days/weeks for a full codec. Consequently, we significantly increased fast-prototyping speed by simultaneously reducing hardware requirements. We verified our findings against full neural codecs. Our findings led to a modification of the straight-through estimator, which stabilized training and improved the state-of-the-art descript-audio-codec. We plan to use the proposed efficient evaluation framework to analyze additional gradient estimators like ReinMax or SPIGOT and investigate the effect of different layers/activations before and after the quantizer on the neural codec. Additionally, we plan to thoroughly examine the modification of the straight-through estimator and its impact on neural codecs. 
\section{Impact Statement}
This paper presents work whose goal is to advance the field of Machine Learning. There are many potential societal consequences of our work, none which we feel must be specifically highlighted here.
\clearpage
\balance
\bibliography{example_paper}
\balance
\bibliographystyle{icml2025}


\end{document}

%% file: Figures/Plot0loss.tex
\begin{tikzpicture}

\definecolor{darkgray176}{RGB}{176,176,176}
\definecolor{darkorange25512714}{RGB}{255,127,14}
\definecolor{lightgray204}{RGB}{204,204,204}
\definecolor{steelblue31119180}{RGB}{31,119,180}

\begin{axis}[
name = plot1,
    width=3cm,  
    height=3cm,  
    scale only axis, 
    xlabel={Epochs},
legend cell align={center},
legend style={
  fill opacity=0.8,
  draw opacity=1,
  text opacity=1,
  at={(1.25,1.5)},
  anchor=north,
  draw=lightgray204,
  legend columns = -1
  mark=*
},
legend image code/.code={
        \draw[mark options={scale=1}, mark=*] plot coordinates {(0.3cm,0cm)}; 
    },
tick align=outside,
tick pos=left,
x grid style={darkgray176},
xmajorgrids,
xmin=0, xmax=100,
xtick style={color=black},
y grid style={darkgray176},
ylabel={},
title={MSE-Loss (\ref{equ:MSELoss})},
ymajorgrids,
ymin=-0.05, 
ymax=0.7,
ytick style={color=black},
]
\addplot [semithick, steelblue31119180]
table {%
0 0.491439277626574
1 0.145839979875833
2 0.040147550188005
3 0.0221479450738989
4 0.00149579956594971
5 0.000185314791179735
6 2.88211019144455e-05
7 9.26099238910183e-06
8 5.63369642819112e-06
9 4.19628487658485e-06
10 3.45550454594701e-06
11 3.07767361245226e-06
12 2.69274032850575e-06
13 2.5199281473185e-06
14 2.32981874470539e-06
15 2.34445380689152e-06
16 2.23870044637131e-06
17 2.18607506899549e-06
18 2.14165844633563e-06
19 2.14033544194869e-06
20 2.19737296721045e-06
21 1.99292602428924e-06
22 2.05227904789707e-06
23 2.06425206110727e-06
24 2.0497238110101e-06
25 2.09024653652873e-06
26 1.96800227211469e-06
27 1.98377456345733e-06
28 2.04983100771627e-06
29 1.93138959105754e-06
30 1.97143960096093e-06
31 1.96656108060764e-06
32 1.96128012035343e-06
33 1.87406848142269e-06
34 1.95899548123064e-06
35 2.06950092014527e-06
36 1.77598596714649e-06
37 1.89258909054116e-06
38 1.86770988570095e-06
39 1.90023664551175e-06
40 1.98853274846311e-06
41 1.73420043834979e-06
42 1.83053138744999e-06
43 1.83887237433022e-06
44 1.84024018024732e-06
45 1.90997533066628e-06
46 1.73806030858443e-06
47 1.80118299307216e-06
48 1.77944000292866e-06
49 1.76767705089595e-06
50 1.78103714961068e-06
51 1.75996178718733e-06
52 1.79876262444878e-06
53 1.68875101220228e-06
54 1.72277751419593e-06
55 1.74380437144303e-06
56 1.65202610490772e-06
57 1.76343048156535e-06
58 1.62815765590901e-06
59 1.71006245526205e-06
60 1.67780812488743e-06
61 1.72043847649969e-06
62 1.65427349096531e-06
63 1.60840335319814e-06
64 1.6785292990398e-06
65 1.62734447482625e-06
66 1.66653455679233e-06
67 1.55285236923791e-06
68 1.59857226366006e-06
69 1.5729478062908e-06
70 1.58914339040814e-06
71 1.55204935845105e-06
72 1.56630324249829e-06
73 1.57998718153763e-06
74 1.52933544721759e-06
75 1.59393987128011e-06
76 1.61161236163954e-06
77 1.4681082392279e-06
78 1.52794630263894e-06
79 1.49592511971763e-06
80 1.48590521772443e-06
81 1.55174371780464e-06
82 1.41478302101003e-06
83 1.49139737778946e-06
84 1.48048299807561e-06
85 1.51897484758891e-06
86 1.38714931934843e-06
87 1.49776836886953e-06
88 1.41997981466642e-06
89 1.47431763927672e-06
90 1.42272246303093e-06
91 1.4149008913151e-06
92 1.39491322596242e-06
93 1.43892097185226e-06
94 1.3585801499705e-06
95 1.36681164034219e-06
96 1.38891166662127e-06
97 1.37259174018019e-06
98 1.36007890725973e-06
99 1.33940678363167e-06
};
\addlegendentry{no Quantizer}
\addplot [semithick, darkorange25512714]
table {%
0 0.549000259801745
1 0.299871038094163
2 0.242189251571894
3 0.207308624267578
4 0.197074625216424
5 0.185634178325534
6 0.182380964674056
7 0.178326630882919
8 0.175201226659119
9 0.173283803023398
10 0.172094418533146
11 0.171343371406198
12 0.170765378408134
13 0.169521951183677
14 0.167103111922741
15 0.157875367105007
16 0.154614733763039
17 0.153081586278975
18 0.150951494216919
19 0.149476350657642
20 0.147199265860021
21 0.144910294152796
22 0.143213532932103
23 0.14128946762532
24 0.137701354525983
25 0.136567906014621
26 0.134132204532623
27 0.132960757054389
28 0.132355139888823
29 0.131965770825744
30 0.131745044507086
31 0.131587471209466
32 0.13147689602524
33 0.131349040202796
34 0.131103035457432
35 0.130517799466848
36 0.130464115925133
37 0.130513427942991
38 0.130506151787937
39 0.130541185207665
40 0.130534353867173
41 0.130492865376174
42 0.130522380650043
43 0.130526529125869
44 0.130550112396479
45 0.130531256340444
46 0.130503811851144
47 0.130515778191388
48 0.130510073944926
49 0.130536538757384
50 0.130526625037193
51 0.130501929484308
52 0.130520934864879
53 0.130509537316859
54 0.130512963078916
55 0.130522072032094
56 0.130513535268605
57 0.130511100739241
58 0.130539780512452
59 0.130513597756624
60 0.130501266375184
61 0.130528594397008
62 0.130528959475458
63 0.130543941237032
64 0.130528621524572
65 0.130521844916046
66 0.130523137196898
67 0.130502953402698
68 0.130502688154578
69 0.130511386707425
70 0.130524727001786
71 0.130521879352629
72 0.130487557888031
73 0.130523864358664
74 0.130489279292524
75 0.130496246568859
76 0.130508339770138
77 0.130523177161813
78 0.130494157969952
79 0.130501418083906
80 0.130529735192657
81 0.130518131420016
82 0.130508684828877
83 0.130500476054847
84 0.130506072625518
85 0.130520362116396
86 0.13052210765332
87 0.130500621564686
88 0.130523995541036
89 0.130495601221919
90 0.130508861780167
91 0.130531919181347
92 0.130510828539729
93 0.130509101711214
94 0.130512753784657
95 0.130492839120328
96 0.130500469848514
97 0.130490967720747
98 0.130495262987912
99 0.130493671834469
};
\addlegendentry{$\text{SQ}^{\text{CL}}_\text{STE}$ }
\end{axis}

\begin{axis}[
        at={(plot1.east)}, 
    anchor=west,
    width=3cm,  
    height=3cm,  
     xshift=1.5cm, 
    scale only axis, 
    title={MA-E (\ref{equ:MAE})},
log basis y={10},
tick align=outside,
tick pos=left,
x grid style={darkgray176},
xlabel={Epochs},
xmajorgrids,
xmin=0, xmax=100,
xtick style={color=black},
y grid style={darkgray176},
ymajorgrids,
ymin=0.37772735089694, ymax=302220.574977491,
ymode=log,
ytick style={color=black}
]
\addplot [semithick, steelblue31119180]
table {%
0 0.967797929048538
1 1.18999158143997
2 1.3280640522639
3 1.45739589532216
4 1.48696814775467
5 1.49455015858014
6 1.4984413087368
7 1.50753791928291
8 1.48351823091507
9 1.49526984492938
10 1.48783353964488
11 1.47498019933701
12 1.48275196949641
13 1.48674787680308
14 1.46704434951146
15 1.4647004087766
16 1.47203040917714
17 1.46624130010605
18 1.47390285730362
19 1.46274871031443
20 1.45827481945356
21 1.4549953798453
22 1.47385632594426
23 1.45014885862668
24 1.4484273036321
25 1.44306247830391
26 1.45101089477539
27 1.46076002319654
28 1.4252666870753
29 1.44686559240023
30 1.42446284294128
31 1.43271813591321
32 1.44545921683311
33 1.43073971271515
34 1.43180764913559
35 1.4184127052625
36 1.42323799530665
37 1.43259494105975
38 1.41359353661537
39 1.40981947978338
40 1.40724654197693
41 1.406930210193
42 1.41545903881391
43 1.41366568605105
44 1.3914620300134
45 1.41088707844416
46 1.41070099075635
47 1.38227074742317
48 1.38994129101435
49 1.38415230512619
50 1.40318608283997
51 1.39011055628459
52 1.39560886224111
53 1.38020410339038
54 1.39084505240122
55 1.38685316840808
56 1.36806154648463
57 1.38207839330037
58 1.3816189746062
59 1.35772135059039
60 1.35888500610987
61 1.36058977842331
62 1.37051539222399
63 1.36771312554677
64 1.36801095207532
65 1.35492933392525
66 1.36689975063006
67 1.35233320991198
68 1.34628750681877
69 1.34210849801699
70 1.34686796466509
71 1.35329723159472
72 1.3503042280674
73 1.33104382157326
74 1.34038360118866
75 1.33892426292102
76 1.324663601319
77 1.33607966899872
78 1.32023986379306
79 1.31750364899635
80 1.32620007594426
81 1.31737558444341
82 1.31540066798528
83 1.32045937975248
84 1.32427805860837
85 1.29337703386943
86 1.32230117917061
87 1.32707501451174
88 1.32695627212524
89 1.30781185030937
90 1.30090575218201
91 1.31222044229507
92 1.3084713101387
93 1.3018348634243
94 1.2810719927152
95 1.27780377070109
96 1.30311262607574
97 1.29654068748156
98 1.288038311402
99 1.29606095949809
};
\addplot [semithick, darkorange25512714]
table {%
0 1.15177705486616
1 1.49898407856623
2 1.68984109958013
3 1.74100794792175
4 1.81043399969737
5 1.80887944698334
6 1.81068992614746
7 1.80273396571477
8 1.79722727139791
9 1.80120892524719
10 1.79876539707184
11 1.8107432325681
12 1.81065675814947
13 1.82434448401133
14 1.8075327595075
15 1.9066148519516
16 1.90987679560979
17 1.90051119724909
18 1.88925451437632
19 1.88267554442088
20 1.87240897019704
21 1.85495265722275
22 1.84392866690954
23 1.8233410914739
24 1.80795597235362
25 1.81095338265101
26 1.7882888674736
27 1.79436893860499
28 1.7830516854922
29 1.78319657246272
30 1.78521246512731
31 1.78399416208267
32 1.77897806962331
33 1.78484630187352
34 1.78965667883555
35 1.78724503119787
36 1.77722928126653
37 1.78671556711197
38 1.78292543490728
39 1.77811615069707
40 1.77549083630244
41 1.78069831530253
42 1.77355553309123
43 1.77956559658051
44 1.78367565870285
45 1.77791599432627
46 1.77011037667592
47 1.7834709127744
48 1.77421322266261
49 1.77529484033585
50 1.78160897095998
51 1.77904099623362
52 1.77556119362513
53 1.78328501383464
54 1.78391941785812
55 1.78135044972102
56 1.7737685362498
57 1.78358612060547
58 1.77882724603017
59 1.7719974120458
60 1.77522218624751
61 1.77480307420095
62 1.77991149822871
63 1.77920944690704
64 1.77348047494888
65 1.78633494377136
66 1.77663913567861
67 1.77468335231145
68 1.78009341160456
69 1.78007904291153
70 1.77263360420863
71 1.77128464778264
72 1.7840326944987
73 1.77595264911652
74 1.77936276594798
75 1.78149652481079
76 1.78397656679153
77 1.77532234191895
78 1.78522520860036
79 1.77818075418472
80 1.78747274080912
81 1.77702915668488
82 1.77857061624527
83 1.77995453675588
84 1.7965615272522
85 1.77666591008504
86 1.78238356908162
87 1.77309301694234
88 1.78007437785467
89 1.77789382537206
90 1.78000446160634
91 1.78402543465296
92 1.78007783095042
93 1.77365928490957
94 1.78022813002268
95 1.77402989466985
96 1.78552823464076
97 1.77882620096207
98 1.78523919582367
99 1.7766180674235
};
\end{axis}
\end{tikzpicture}

%% file: Figures/Plot1loss.tex
\begin{tikzpicture}

\definecolor{crimson2143940}{RGB}{214,39,40}
\definecolor{darkgray176}{RGB}{176,176,176}
\definecolor{darkorange25512714}{RGB}{255,127,14}
\definecolor{forestgreen4416044}{RGB}{44,160,44}
\definecolor{lightgray204}{RGB}{204,204,204}
\definecolor{steelblue31119180}{RGB}{31,119,180}

\begin{axis}[
name = plot1,
    width=3cm,  
    height=3cm,  
    scale only axis, 
    xlabel={Epochs},
legend cell align={center},
legend style={
  fill opacity=0.8,
  draw opacity=1,
  text opacity=1,
  at={(1.25,1.5)},
  anchor=north,
  draw=lightgray204,
  legend columns = -1
  mark=*
},
legend image code/.code={
        \draw[mark options={scale=1}, mark=*] plot coordinates {(0.3cm,0cm)}; 
    },
tick align=outside,
tick pos=left,
x grid style={darkgray176},
xmajorgrids,
xmin=0, xmax=100,
xtick style={color=black},
y grid style={darkgray176},
ylabel={},
title={MSE-Loss (\ref{equ:MSELoss})},
ymajorgrids,
ymin=0, 
ymax=0.7,
ytick style={color=black},
]
\addplot [semithick, steelblue31119180]
table {%
0 0.611941409394145
1 0.404460925444961
2 0.355163639776409
3 0.339247392460704
4 0.324812070466578
5 0.301795873902738
6 0.292822175815701
7 0.290735897243023
8 0.292805431999266
9 0.290305355966091
10 0.291941600158811
11 0.288333343707025
12 0.288737422563136
13 0.296677809603512
14 0.292281404785812
15 0.290858758054674
16 0.290767313160002
17 0.292714920699596
18 0.290587794758379
19 0.289766863293946
20 0.288406750351191
21 0.293193183898926
22 0.291701209932566
23 0.290007071204483
24 0.292503394111991
25 0.293332264944911
26 0.291972432099283
27 0.291046844929457
28 0.287716829553247
29 0.291602838911116
30 0.284840544752777
31 0.280512229338288
32 0.280809077531099
33 0.277989823311567
34 0.278361989110708
35 0.279393471904099
36 0.279003665581346
37 0.279020595915616
38 0.278237650796771
39 0.277349738404155
40 0.277278342619538
41 0.273919279158115
42 0.279123209051788
43 0.27709675514698
44 0.28020255998522
45 0.27470947098732
46 0.280286003962159
47 0.27813413977623
48 0.278169064290822
49 0.279392674587667
50 0.277171799391508
51 0.276486691772938
52 0.277413161180913
53 0.279150147378445
54 0.281031813181937
55 0.275608771234751
56 0.27366885086149
57 0.279172012083232
58 0.278515524618328
59 0.276100231125951
60 0.278760932698846
61 0.275043711975217
62 0.279179656289518
63 0.278528353877366
64 0.276965564243495
65 0.277563843168318
66 0.275165799811482
67 0.2789791867733
68 0.276749472655356
69 0.276790994957089
70 0.276459479801357
71 0.276327098228037
72 0.279078826345503
73 0.27826146197319
74 0.278874768920243
75 0.277815463848412
76 0.278804229311645
77 0.275874389119446
78 0.278570503793657
79 0.275502957567573
80 0.278709277048707
81 0.277353476390243
82 0.280052238218486
83 0.279736127734184
84 0.278723088487983
85 0.278621223472059
86 0.279907686054707
87 0.275282921276987
88 0.273415324039757
89 0.274746660880744
90 0.274893422745168
91 0.277455612197518
92 0.274458181723952
93 0.277263621002436
94 0.278553314588964
95 0.281348010055721
96 0.279461028672755
97 0.277654385901988
98 0.279517267264426
99 0.276829372189939
};
\addlegendentry{NA}
\addplot [semithick, darkorange25512714]
table {%
0 0.608796063810587
1 0.404834068283439
2 0.353451243817806
3 0.338568493880332
4 0.323459391139448
5 0.311533840574324
6 0.300045564420521
7 0.291719058051705
8 0.292339117482305
9 0.291288265392184
10 0.290410711705685
11 0.290524671278894
12 0.293199801497161
13 0.293025554820895
14 0.290531463526189
15 0.295489782825112
16 0.292294562637806
17 0.292943701155484
18 0.290817392267287
19 0.293005076251924
20 0.294164174884558
21 0.291099690787494
22 0.292309767723083
23 0.288032758355141
24 0.286823302768171
25 0.293853967666626
26 0.294363110654056
27 0.292947210237384
28 0.294658658236265
29 0.293523136638105
30 0.293943503648043
31 0.289424084164202
32 0.295608305335045
33 0.287112329237163
34 0.291889309346676
35 0.289596081763506
36 0.290031305097044
37 0.289163750000298
38 0.289397771321237
39 0.277202979832888
40 0.278286978907883
41 0.274547535397112
42 0.275911723800004
43 0.277363488964736
44 0.277568548239768
45 0.273078170023859
46 0.278599904976785
47 0.278825132340193
48 0.279550102636218
49 0.278170816399157
50 0.277482765451074
51 0.276931995548308
52 0.27972166389972
53 0.278546011075377
54 0.278506645992398
55 0.282398849375546
56 0.275060715161264
57 0.275370138250291
58 0.279895563319325
59 0.277708165705204
60 0.281192395374179
61 0.27682488746196
62 0.275595603443682
63 0.277571296684444
64 0.280766646973789
65 0.27943359337002
66 0.278589633107185
67 0.275965581551194
68 0.280830531902611
69 0.273645771220326
70 0.278702991873026
71 0.274352182485163
72 0.273271102510393
73 0.27945484072715
74 0.276404253445566
75 0.277227476261556
76 0.273234584100544
77 0.278205651275814
78 0.276227233588696
79 0.279279652908444
80 0.276720661975443
81 0.273039626404643
82 0.276899457156658
83 0.277564294412732
84 0.274600497633219
85 0.27512222328037
86 0.280471811644733
87 0.286434203542769
88 0.275768381729722
89 0.283131299182773
90 0.277525585323572
91 0.276315912701189
92 0.279433982238174
93 0.277927031427622
94 0.276872227557004
95 0.275729609794915
96 0.278945564739406
97 0.276855052880943
98 0.278311800837517
99 0.279406821645796
};
\addlegendentry{$\text{NA}^\text{CL}$}
\addplot [semithick, forestgreen4416044]
table {%
0 0.55088772341609
1 0.364599603518844
2 0.368866781711578
3 0.377288254261017
4 0.388566744953394
5 0.401835393205285
6 0.414005573749542
7 0.424335187733173
8 0.433307941555977
9 0.440919484496117
10 0.447635944843292
11 0.453682038679719
12 0.458860293194652
13 0.463521628290415
14 0.467869847580791
15 0.471703230798244
16 0.475401790723205
17 0.478585365355015
18 0.481512870833278
19 0.484425481408834
20 0.487144548967481
21 0.489412767782807
22 0.491777189701796
23 0.493936263769865
24 0.495904034361243
25 0.497762332350016
26 0.499550337582827
27 0.501130907699466
28 0.502885586768389
29 0.504318796187639
30 0.505741798982024
31 0.507270608186722
32 0.508436697110534
33 0.509720933899283
34 0.510865337014198
35 0.511979941368103
36 0.513019118696451
37 0.513959438443184
38 0.514955953866243
39 0.516012669444084
40 0.516867266148329
41 0.517744843870401
42 0.51857935076952
43 0.519320396065712
44 0.519961907505989
45 0.520658597141504
46 0.521395146280527
47 0.521903126507998
48 0.522700056582689
49 0.52323027163744
50 0.523826842874289
51 0.524266020983458
52 0.524801029682159
53 0.525233933061361
54 0.525834440082312
55 0.526259167104959
56 0.526617806315422
57 0.527145440518856
58 0.527537268251181
59 0.527885058254004
60 0.528327745348215
61 0.52878401529789
62 0.529222670078278
63 0.529591565400362
64 0.529899104863405
65 0.530336342602968
66 0.530688235938549
67 0.530990009218454
68 0.531487732738256
69 0.531842892467976
70 0.532275775909424
71 0.532542983800173
72 0.532737836956978
73 0.533120610415936
74 0.533493458986282
75 0.533814318478107
76 0.534058399617672
77 0.534519419878721
78 0.534666323184967
79 0.534989660829306
80 0.535261952996254
81 0.535537146002054
82 0.535684410214424
83 0.535837888389826
84 0.536177230507135
85 0.53638651779294
86 0.536517620921135
87 0.536831999242306
88 0.536970202863216
89 0.537155061453581
90 0.537432105869055
91 0.53773717880249
92 0.53779073548317
93 0.538084347188473
94 0.538191051691771
95 0.53855269575119
96 0.53880144482851
97 0.538929655462503
98 0.5389938133955
99 0.539319223552942
};
\addlegendentry{$SQ_\text{STE}$}
\addplot [semithick, crimson2143940]
table {%
0 0.549000259801745
1 0.299871038094163
2 0.242189251571894
3 0.207308624267578
4 0.197074625216424
5 0.185634178325534
6 0.182380964674056
7 0.178326630882919
8 0.175201226659119
9 0.173283803023398
10 0.172094418533146
11 0.171343371406198
12 0.170765378408134
13 0.169521951183677
14 0.167103111922741
15 0.157875367105007
16 0.154614733763039
17 0.153081586278975
18 0.150951494216919
19 0.149476350657642
20 0.147199265860021
21 0.144910294152796
22 0.143213532932103
23 0.14128946762532
24 0.137701354525983
25 0.136567906014621
26 0.134132204532623
27 0.132960757054389
28 0.132355139888823
29 0.131965770825744
30 0.131745044507086
31 0.131587471209466
32 0.13147689602524
33 0.131349040202796
34 0.131103035457432
35 0.130517799466848
36 0.130464115925133
37 0.130513427942991
38 0.130506151787937
39 0.130541185207665
40 0.130534353867173
41 0.130492865376174
42 0.130522380650043
43 0.130526529125869
44 0.130550112396479
45 0.130531256340444
46 0.130503811851144
47 0.130515778191388
48 0.130510073944926
49 0.130536538757384
50 0.130526625037193
51 0.130501929484308
52 0.130520934864879
53 0.130509537316859
54 0.130512963078916
55 0.130522072032094
56 0.130513535268605
57 0.130511100739241
58 0.130539780512452
59 0.130513597756624
60 0.130501266375184
61 0.130528594397008
62 0.130528959475458
63 0.130543941237032
64 0.130528621524572
65 0.130521844916046
66 0.130523137196898
67 0.130502953402698
68 0.130502688154578
69 0.130511386707425
70 0.130524727001786
71 0.130521879352629
72 0.130487557888031
73 0.130523864358664
74 0.130489279292524
75 0.130496246568859
76 0.130508339770138
77 0.130523177161813
78 0.130494157969952
79 0.130501418083906
80 0.130529735192657
81 0.130518131420016
82 0.130508684828877
83 0.130500476054847
84 0.130506072625518
85 0.130520362116396
86 0.13052210765332
87 0.130500621564686
88 0.130523995541036
89 0.130495601221919
90 0.130508861780167
91 0.130531919181347
92 0.130510828539729
93 0.130509101711214
94 0.130512753784657
95 0.130492839120328
96 0.130500469848514
97 0.130490967720747
98 0.130495262987912
99 0.130493671834469
};
\addlegendentry{$SQ_\text{STE}^\text{CL}$}
\end{axis}

\begin{axis}[
        at={(plot1.east)}, 
    anchor=west,
    width=3cm,  
    height=3cm,  
     xshift=1.5cm, 
    scale only axis, 
    title={MA-E (\ref{equ:MAE})},
log basis y={10},
tick align=outside,
tick pos=left,
x grid style={darkgray176},
xlabel={Epochs},
xmajorgrids,
xmin=0, xmax=100,
xtick style={color=black},
y grid style={darkgray176},
ymajorgrids,
ymin=0.37772735089694, ymax=302220.574977491,
ymode=log,
ytick style={color=black}
]
\addplot [semithick, steelblue31119180]
table {%
0 0.784316565593084
1 0.781134277582169
2 0.75108015537262
3 0.747364002466202
4 0.740072242418925
5 0.742147266864777
6 0.734453016519547
7 0.725739912192027
8 0.725482821464539
9 0.725872256358465
10 0.71623806754748
11 0.717047673463821
12 0.724361884593964
13 0.710147565603256
14 0.71484912832578
15 0.710747335354487
16 0.718673286835353
17 0.720987143119176
18 0.712739396095276
19 0.71228361527125
20 0.712281819184621
21 0.708043175935745
22 0.713457111517588
23 0.707725787162781
24 0.710828079779943
25 0.710757700602214
26 0.708714254697164
27 0.707544475793839
28 0.710333482424418
29 0.71262610356013
30 0.712401813268661
31 0.718252156178157
32 0.716661332050959
33 0.71257671713829
34 0.714103603363037
35 0.715125119686127
36 0.717597754796346
37 0.71170742114385
38 0.715586141745249
39 0.713813449939092
40 0.716613757610321
41 0.713593037923177
42 0.712899752457937
43 0.716273045539856
44 0.716301995515823
45 0.71634173989296
46 0.715151470899582
47 0.720994468530019
48 0.719423288106918
49 0.709180527925491
50 0.714239386717478
51 0.714714256922404
52 0.716577911376953
53 0.715655094385147
54 0.718004085620244
55 0.715379571914673
56 0.718934539953868
57 0.715375073750814
58 0.7126762231191
59 0.718821219603221
60 0.718017041683197
61 0.718857727448146
62 0.714162311951319
63 0.712864913543065
64 0.719291679064433
65 0.722002234061559
66 0.721423623959223
67 0.712879719336828
68 0.719574244817098
69 0.722789339224498
70 0.716404211521149
71 0.719432612260183
72 0.722413490215937
73 0.71859184106191
74 0.722885553042094
75 0.721519402662913
76 0.712608919541041
77 0.720148783922195
78 0.72747211654981
79 0.716309491793315
80 0.723044633865356
81 0.720361840724945
82 0.716077470779419
83 0.715186387300491
84 0.727467252810796
85 0.722101583083471
86 0.71787742972374
87 0.724995827674866
88 0.730283008019129
89 0.718816594282786
90 0.718360368410746
91 0.722093685468038
92 0.71898397008578
93 0.715825537840525
94 0.726787326733271
95 0.721089694897334
96 0.722597585121791
97 0.723732564846675
98 0.724596387147903
99 0.73590941230456
};
\addplot [semithick, darkorange25512714]
table {%
0 0.780062007904053
1 0.785003300507863
2 0.760595844189326
3 0.743645322322845
4 0.738197988271713
5 0.727404369910558
6 0.73832106590271
7 0.720802754163742
8 0.72067355910937
9 0.713183629512787
10 0.71515442331632
11 0.705765650669734
12 0.712433300415675
13 0.712155882517497
14 0.704875973860423
15 0.708641225099564
16 0.70104883313179
17 0.708326105276744
18 0.701419488588969
19 0.70841281414032
20 0.702454942464829
21 0.704174659649531
22 0.704577163855235
23 0.717927555243174
24 0.71203598578771
25 0.70237233042717
26 0.705128888289134
27 0.707403345902761
28 0.706829740603765
29 0.703698541720708
30 0.700653398036957
31 0.706547838449478
32 0.703005766868591
33 0.707197372118632
34 0.701774646838506
35 0.702649648984273
36 0.705712916453679
37 0.705201524496079
38 0.706422601143519
39 0.714035723606745
40 0.718191597859065
41 0.714887227614721
42 0.710324345032374
43 0.705266364415487
44 0.713340729475021
45 0.710121959447861
46 0.713599592447281
47 0.704361832141876
48 0.703318081299464
49 0.707208555936813
50 0.708930063247681
51 0.708541176716487
52 0.706024523576101
53 0.706982847054799
54 0.713527637720108
55 0.712104618549347
56 0.715284230311712
57 0.714014929533005
58 0.718396516640981
59 0.712120225032171
60 0.70300657749176
61 0.712330993016561
62 0.705399974187215
63 0.716652369499207
64 0.711654380957286
65 0.715257356564204
66 0.709045793612798
67 0.710406134525935
68 0.713286699851354
69 0.713970551888148
70 0.714134921630224
71 0.708393577734629
72 0.71591382821401
73 0.708847548564275
74 0.716458161671956
75 0.717960963646571
76 0.716552470127741
77 0.71113104224205
78 0.718919990460078
79 0.707574218511581
80 0.714642171065013
81 0.714416901270548
82 0.706916064023972
83 0.711730722586314
84 0.718201941251755
85 0.708433689673742
86 0.716953217983246
87 0.710274974505107
88 0.712638219197591
89 0.712405643860499
90 0.717525462309519
91 0.716002144416173
92 0.711754639943441
93 0.713744324445724
94 0.711945599317551
95 0.720099818706512
96 0.712131476402283
97 0.712278308471044
98 0.713869651158651
99 0.712531715631485
};
\addplot [semithick, forestgreen4416044]
table {%
0 1.52191224098206
1 4.62942314147949
2 10.7700797080994
3 20.8528925577799
4 36.3747828801473
5 58.6029336293538
6 88.5621025085449
7 125.331676737467
8 172.717534383138
9 231.091755167643
10 297.690074666341
11 374.748385620117
12 470.715333048503
13 584.981895955404
14 705.257332356771
15 840.939318847656
16 996.562082926432
17 1164.62107543945
18 1361.14012451172
19 1563.32934977214
20 1795.80465087891
21 2030.06355794271
22 2333.39939778646
23 2611.14085286458
24 2954.9336710612
25 3288.2523844401
26 3671.10323079427
27 4043.10830891927
28 4452.31261393229
29 4877.00172526042
30 5379.12698567708
31 5833.47943522135
32 6415.72179361979
33 7065.05356445312
34 7571.12750651042
35 8357.25302734375
36 8967.70460611979
37 9765.99383138021
38 10491.2435546875
39 11156.7108072917
40 12068.3065917969
41 12973.3904459635
42 13800.2222981771
43 14856.6091145833
44 15786.3408854167
45 16714.0112630208
46 17757.7693684896
47 18953.1794596354
48 20258.6350585938
49 21322.02734375
50 22663.4591471354
51 23886.9681640625
52 25237.6015299479
53 26856.3376953125
54 28415.2879557292
55 29912.9180338542
56 31284.2891276042
57 32826.7267578125
58 34398.5539713542
59 36384.1580729167
60 38069.0947265625
61 39918.037890625
62 42019.1438151042
63 44091.3958984375
64 45745.9811848958
65 48158.7407552083
66 50503.2204427083
67 52731.0869791667
68 54221.2240885417
69 57230.4666666667
70 60084.062109375
71 62508.45625
72 65623.8298177083
73 66958.82265625
74 69791.838671875
75 72859.9966145833
76 76534.0283854167
77 78393.4149739583
78 80455.669921875
79 84537.4856770833
80 88595.2701822917
81 90587.526953125
82 93212.7028645833
83 98404.0579427083
84 100310.597395833
85 104992.432682292
86 107600.389583333
87 111554.88125
88 115896.441666667
89 120097.72578125
90 123141.6984375
91 127448.078385417
92 129743.982552083
93 134410.966927083
94 140104.898958333
95 143944.02421875
96 148692.367708333
97 153510.48828125
98 157471.023958333
99 162929.313541667
};
\addplot [semithick, crimson2143940]
table {%
0 1.15177705486616
1 1.49898407856623
2 1.68984109958013
3 1.74100794792175
4 1.81043399969737
5 1.80887944698334
6 1.81068992614746
7 1.80273396571477
8 1.79722727139791
9 1.80120892524719
10 1.79876539707184
11 1.8107432325681
12 1.81065675814947
13 1.82434448401133
14 1.8075327595075
15 1.9066148519516
16 1.90987679560979
17 1.90051119724909
18 1.88925451437632
19 1.88267554442088
20 1.87240897019704
21 1.85495265722275
22 1.84392866690954
23 1.8233410914739
24 1.80795597235362
25 1.81095338265101
26 1.7882888674736
27 1.79436893860499
28 1.7830516854922
29 1.78319657246272
30 1.78521246512731
31 1.78399416208267
32 1.77897806962331
33 1.78484630187352
34 1.78965667883555
35 1.78724503119787
36 1.77722928126653
37 1.78671556711197
38 1.78292543490728
39 1.77811615069707
40 1.77549083630244
41 1.78069831530253
42 1.77355553309123
43 1.77956559658051
44 1.78367565870285
45 1.77791599432627
46 1.77011037667592
47 1.7834709127744
48 1.77421322266261
49 1.77529484033585
50 1.78160897095998
51 1.77904099623362
52 1.77556119362513
53 1.78328501383464
54 1.78391941785812
55 1.78135044972102
56 1.7737685362498
57 1.78358612060547
58 1.77882724603017
59 1.7719974120458
60 1.77522218624751
61 1.77480307420095
62 1.77991149822871
63 1.77920944690704
64 1.77348047494888
65 1.78633494377136
66 1.77663913567861
67 1.77468335231145
68 1.78009341160456
69 1.78007904291153
70 1.77263360420863
71 1.77128464778264
72 1.7840326944987
73 1.77595264911652
74 1.77936276594798
75 1.78149652481079
76 1.78397656679153
77 1.77532234191895
78 1.78522520860036
79 1.77818075418472
80 1.78747274080912
81 1.77702915668488
82 1.77857061624527
83 1.77995453675588
84 1.7965615272522
85 1.77666591008504
86 1.78238356908162
87 1.77309301694234
88 1.78007437785467
89 1.77789382537206
90 1.78000446160634
91 1.78402543465296
92 1.78007783095042
93 1.77365928490957
94 1.78022813002268
95 1.77402989466985
96 1.78552823464076
97 1.77882620096207
98 1.78523919582367
99 1.7766180674235
};

\end{axis}

\end{tikzpicture}

%% file: Figures/Plot2loss.tex
\begin{tikzpicture}

\definecolor{crimson2143940}{RGB}{214,39,40}
\definecolor{darkgray176}{RGB}{176,176,176}
\definecolor{darkorange25512714}{RGB}{255,127,14}
\definecolor{forestgreen4416044}{RGB}{44,160,44}
\definecolor{lightgray204}{RGB}{204,204,204}
\definecolor{steelblue31119180}{RGB}{31,119,180}

\begin{axis}[
name = plot1,
    width=3cm,  
    height=3cm,  
    scale only axis, 
    xlabel={Epochs},
legend cell align={center},
legend style={
  fill opacity=0.8,
  draw opacity=1,
  text opacity=1,
  at={(1.25,1.5)},
  anchor=north,
  draw=lightgray204,
  legend columns = -1
  mark=*
},
legend image code/.code={
        \draw[mark options={scale=1}, mark=*] plot coordinates {(0.3cm,0cm)}; 
    },
tick align=outside,
tick pos=left,
x grid style={darkgray176},
xmajorgrids,
xmin=0, xmax=100,
xtick style={color=black},
y grid style={darkgray176},
ylabel={},
title={MSE-Loss (\ref{equ:MSELoss})},
ymajorgrids,
ymin=0.2, 
ymax=1,
ytick style={color=black},
]
\addplot [semithick, steelblue31119180]
table {%
0 0.611941409394145
1 0.404460925444961
2 0.355163639776409
3 0.339247392460704
4 0.324812070466578
5 0.301795873902738
6 0.292822175815701
7 0.290735897243023
8 0.292805431999266
9 0.290305355966091
10 0.291941600158811
11 0.288333343707025
12 0.288737422563136
13 0.296677809603512
14 0.292281404785812
15 0.290858758054674
16 0.290767313160002
17 0.292714920699596
18 0.290587794758379
19 0.289766863293946
20 0.288406750351191
21 0.293193183898926
22 0.291701209932566
23 0.290007071204483
24 0.292503394111991
25 0.293332264944911
26 0.291972432099283
27 0.291046844929457
28 0.287716829553247
29 0.291602838911116
30 0.284840544752777
31 0.280512229338288
32 0.280809077531099
33 0.277989823311567
34 0.278361989110708
35 0.279393471904099
36 0.279003665581346
37 0.279020595915616
38 0.278237650796771
39 0.277349738404155
40 0.277278342619538
41 0.273919279158115
42 0.279123209051788
43 0.27709675514698
44 0.28020255998522
45 0.27470947098732
46 0.280286003962159
47 0.27813413977623
48 0.278169064290822
49 0.279392674587667
50 0.277171799391508
51 0.276486691772938
52 0.277413161180913
53 0.279150147378445
54 0.281031813181937
55 0.275608771234751
56 0.27366885086149
57 0.279172012083232
58 0.278515524618328
59 0.276100231125951
60 0.278760932698846
61 0.275043711975217
62 0.279179656289518
63 0.278528353877366
64 0.276965564243495
65 0.277563843168318
66 0.275165799811482
67 0.2789791867733
68 0.276749472655356
69 0.276790994957089
70 0.276459479801357
71 0.276327098228037
72 0.279078826345503
73 0.27826146197319
74 0.278874768920243
75 0.277815463848412
76 0.278804229311645
77 0.275874389119446
78 0.278570503793657
79 0.275502957567573
80 0.278709277048707
81 0.277353476390243
82 0.280052238218486
83 0.279736127734184
84 0.278723088487983
85 0.278621223472059
86 0.279907686054707
87 0.275282921276987
88 0.273415324039757
89 0.274746660880744
90 0.274893422745168
91 0.277455612197518
92 0.274458181723952
93 0.277263621002436
94 0.278553314588964
95 0.281348010055721
96 0.279461028672755
97 0.277654385901988
98 0.279517267264426
99 0.276829372189939
};
\addlegendentry{$\text{NA}$}
\addplot [semithick, darkorange25512714]
table {%
0 0.608796063810587
1 0.404834068283439
2 0.353451243817806
3 0.338568493880332
4 0.323459391139448
5 0.311533840574324
6 0.300045564420521
7 0.291719058051705
8 0.292339117482305
9 0.291288265392184
10 0.290410711705685
11 0.290524671278894
12 0.293199801497161
13 0.293025554820895
14 0.290531463526189
15 0.295489782825112
16 0.292294562637806
17 0.292943701155484
18 0.290817392267287
19 0.293005076251924
20 0.294164174884558
21 0.291099690787494
22 0.292309767723083
23 0.288032758355141
24 0.286823302768171
25 0.293853967666626
26 0.294363110654056
27 0.292947210237384
28 0.294658658236265
29 0.293523136638105
30 0.293943503648043
31 0.289424084164202
32 0.295608305335045
33 0.287112329237163
34 0.291889309346676
35 0.289596081763506
36 0.290031305097044
37 0.289163750000298
38 0.289397771321237
39 0.277202979832888
40 0.278286978907883
41 0.274547535397112
42 0.275911723800004
43 0.277363488964736
44 0.277568548239768
45 0.273078170023859
46 0.278599904976785
47 0.278825132340193
48 0.279550102636218
49 0.278170816399157
50 0.277482765451074
51 0.276931995548308
52 0.27972166389972
53 0.278546011075377
54 0.278506645992398
55 0.282398849375546
56 0.275060715161264
57 0.275370138250291
58 0.279895563319325
59 0.277708165705204
60 0.281192395374179
61 0.27682488746196
62 0.275595603443682
63 0.277571296684444
64 0.280766646973789
65 0.27943359337002
66 0.278589633107185
67 0.275965581551194
68 0.280830531902611
69 0.273645771220326
70 0.278702991873026
71 0.274352182485163
72 0.273271102510393
73 0.27945484072715
74 0.276404253445566
75 0.277227476261556
76 0.273234584100544
77 0.278205651275814
78 0.276227233588696
79 0.279279652908444
80 0.276720661975443
81 0.273039626404643
82 0.276899457156658
83 0.277564294412732
84 0.274600497633219
85 0.27512222328037
86 0.280471811644733
87 0.286434203542769
88 0.275768381729722
89 0.283131299182773
90 0.277525585323572
91 0.276315912701189
92 0.279433982238174
93 0.277927031427622
94 0.276872227557004
95 0.275729609794915
96 0.278945564739406
97 0.276855052880943
98 0.278311800837517
99 0.279406821645796
};
\addlegendentry{$\text{NA}^\text{CL}$}
\addplot [semithick, forestgreen4416044]
table {%
0 0.611609450370073
1 0.439915159359574
2 0.402905011504889
3 0.383407583221793
4 0.378339802965522
5 0.382299789071083
6 0.384070707291365
7 0.389385624021292
8 0.399069177210331
9 0.404827949911356
10 0.412929787904024
11 0.414823229983449
12 0.421921906873584
13 0.425825444385409
14 0.43003033643961
15 0.433517695590854
16 0.438393251121044
17 0.437865211963654
18 0.439733000442386
19 0.448825848251581
20 0.44746831817925
21 0.451768790438771
22 0.456774653479457
23 0.459475716635585
24 0.460669831678271
25 0.467696123689413
26 0.470393957212567
27 0.47330911564827
28 0.479266878798604
29 0.492035465598106
30 0.495099705278873
31 0.506255769014359
32 0.501853556618094
33 0.504647214472294
34 0.507795336171985
35 0.517213715299964
36 0.527184344947338
37 0.540708477377892
38 0.535105723246932
39 0.561149383425713
40 0.57866429579258
41 0.53783510453999
42 0.590163114398718
43 0.557026644676924
44 0.61548131018877
45 0.54224068044126
46 0.597406392633915
47 0.700817015364766
48 0.739486065238714
49 0.64181193523109
50 0.59636370061338
51 0.614021787241101
52 0.735419025689363
53 0.590475044742227
54 0.652752056062222
55 0.664430199787021
56 0.720020437479019
57 0.646270394340158
58 0.691383462801576
59 0.618400651171803
60 0.667103306367993
61 0.683399421453476
62 0.671754186838865
63 0.655481513738632
64 0.642903120398521
65 0.645933763623238
66 0.849703469723463
67 0.634023216992617
68 0.706610255494714
69 0.70335243485868
70 0.701880310982466
71 0.743811122655869
72 0.694135029777884
73 0.734894439533353
74 0.683351158410311
75 0.729692776799202
76 0.737865243852138
77 0.656588634103537
78 0.746453127741814
79 0.786057748436928
80 0.769413703620434
81 0.746070098757744
82 0.729212529271841
83 0.773686207860708
84 0.720844190388918
85 0.736456541448832
86 0.730318913727999
87 0.695238190889359
88 0.747054092675447
89 0.69227656596899
90 0.725485775560141
91 0.730593508601189
92 0.687429808497429
93 0.753641017556191
94 0.709013304740191
95 0.688680979758501
96 0.741845514267683
97 0.700211458534002
98 0.665500582426786
99 0.710758947059512
};
\addlegendentry{$\text{NA}_\text{det}$}
\addplot [semithick, crimson2143940]
table {%
0 0.61643166154623
1 0.451805842861533
2 0.409961090907454
3 0.392075737953186
4 0.385575036212802
5 0.382028292134404
6 0.381157822176814
7 0.381767434999347
8 0.385939629092813
9 0.388244169682264
10 0.394128766864538
11 0.393550968199968
12 0.392839833676815
13 0.396085587128997
14 0.400960114747286
15 0.401235269933939
16 0.404315443471074
17 0.40736368817091
18 0.410646104961634
19 0.407733978703618
20 0.410043139457703
21 0.41177126955986
22 0.412383782222867
23 0.413597009524703
24 0.417239158540964
25 0.417437488898635
26 0.415505234986544
27 0.417706749945879
28 0.417414122521877
29 0.421165493309498
30 0.420716576308012
31 0.420093193635345
32 0.420569443538785
33 0.422663229003549
34 0.421872788339853
35 0.426678671315312
36 0.426079013392329
37 0.425104559883475
38 0.42722222545743
39 0.429121752738953
40 0.42655730958283
41 0.427006248027086
42 0.429205042302608
43 0.430995577126741
44 0.429385351166129
45 0.430380379512906
46 0.432773048639297
47 0.431285184592009
48 0.431893689617515
49 0.432349317163229
50 0.430300185695291
51 0.431707549676299
52 0.431425787881017
53 0.433075299277902
54 0.431460279360414
55 0.435230349957943
56 0.435780043512583
57 0.433873699858785
58 0.436460872322321
59 0.436038340106606
60 0.436940508961678
61 0.435898477628827
62 0.435486101523042
63 0.436774514377117
64 0.436168460130692
65 0.439455194503069
66 0.43727204246819
67 0.438427097216249
68 0.440966750845313
69 0.437848162785172
70 0.440606507360935
71 0.440070451557636
72 0.439671826213598
73 0.437487634658813
74 0.441459807395935
75 0.441003379151225
76 0.439396331682801
77 0.442133032754064
78 0.44050989998877
79 0.441478393092751
80 0.443431491866708
81 0.441017071440816
82 0.442220961645246
83 0.443517910316586
84 0.445675541892648
85 0.441425472855568
86 0.446285691589117
87 0.444073230117559
88 0.442379747524858
89 0.444188039451838
90 0.445629195898771
91 0.442071973070502
92 0.445623260036111
93 0.443225308865309
94 0.445463816702366
95 0.443319242119789
96 0.443458917245269
97 0.445811537995934
98 0.446655088037252
99 0.444551042348146
};
\addlegendentry{$\text{NA}^\text{CL}_\text{det}$}
\end{axis}

\begin{axis}[
        at={(plot1.east)}, 
    anchor=west,
    width=3cm,  
    height=3cm,  
     xshift=1.1cm, 
    scale only axis, 
     title={MA-E (\ref{equ:MAE})},
log basis y={10},
tick align=outside,
tick pos=left,
x grid style={darkgray176},
xlabel={Epochs},
xmajorgrids,
xmin=0, xmax=100,
xtick style={color=black},
y grid style={darkgray176},
ymajorgrids,
ymin=0.37772735089694, ymax=302220.574977491,
ymode=log,
ytick style={color=black}
]
\addplot [semithick, steelblue31119180]
table {%
0 0.784316565593084
1 0.781134277582169
2 0.75108015537262
3 0.747364002466202
4 0.740072242418925
5 0.742147266864777
6 0.734453016519547
7 0.725739912192027
8 0.725482821464539
9 0.725872256358465
10 0.71623806754748
11 0.717047673463821
12 0.724361884593964
13 0.710147565603256
14 0.71484912832578
15 0.710747335354487
16 0.718673286835353
17 0.720987143119176
18 0.712739396095276
19 0.71228361527125
20 0.712281819184621
21 0.708043175935745
22 0.713457111517588
23 0.707725787162781
24 0.710828079779943
25 0.710757700602214
26 0.708714254697164
27 0.707544475793839
28 0.710333482424418
29 0.71262610356013
30 0.712401813268661
31 0.718252156178157
32 0.716661332050959
33 0.71257671713829
34 0.714103603363037
35 0.715125119686127
36 0.717597754796346
37 0.71170742114385
38 0.715586141745249
39 0.713813449939092
40 0.716613757610321
41 0.713593037923177
42 0.712899752457937
43 0.716273045539856
44 0.716301995515823
45 0.71634173989296
46 0.715151470899582
47 0.720994468530019
48 0.719423288106918
49 0.709180527925491
50 0.714239386717478
51 0.714714256922404
52 0.716577911376953
53 0.715655094385147
54 0.718004085620244
55 0.715379571914673
56 0.718934539953868
57 0.715375073750814
58 0.7126762231191
59 0.718821219603221
60 0.718017041683197
61 0.718857727448146
62 0.714162311951319
63 0.712864913543065
64 0.719291679064433
65 0.722002234061559
66 0.721423623959223
67 0.712879719336828
68 0.719574244817098
69 0.722789339224498
70 0.716404211521149
71 0.719432612260183
72 0.722413490215937
73 0.71859184106191
74 0.722885553042094
75 0.721519402662913
76 0.712608919541041
77 0.720148783922195
78 0.72747211654981
79 0.716309491793315
80 0.723044633865356
81 0.720361840724945
82 0.716077470779419
83 0.715186387300491
84 0.727467252810796
85 0.722101583083471
86 0.71787742972374
87 0.724995827674866
88 0.730283008019129
89 0.718816594282786
90 0.718360368410746
91 0.722093685468038
92 0.71898397008578
93 0.715825537840525
94 0.726787326733271
95 0.721089694897334
96 0.722597585121791
97 0.723732564846675
98 0.724596387147903
99 0.73590941230456
};
\addplot [semithick, darkorange25512714]
table {%
0 0.780062007904053
1 0.785003300507863
2 0.760595844189326
3 0.743645322322845
4 0.738197988271713
5 0.727404369910558
6 0.73832106590271
7 0.720802754163742
8 0.72067355910937
9 0.713183629512787
10 0.71515442331632
11 0.705765650669734
12 0.712433300415675
13 0.712155882517497
14 0.704875973860423
15 0.708641225099564
16 0.70104883313179
17 0.708326105276744
18 0.701419488588969
19 0.70841281414032
20 0.702454942464829
21 0.704174659649531
22 0.704577163855235
23 0.717927555243174
24 0.71203598578771
25 0.70237233042717
26 0.705128888289134
27 0.707403345902761
28 0.706829740603765
29 0.703698541720708
30 0.700653398036957
31 0.706547838449478
32 0.703005766868591
33 0.707197372118632
34 0.701774646838506
35 0.702649648984273
36 0.705712916453679
37 0.705201524496079
38 0.706422601143519
39 0.714035723606745
40 0.718191597859065
41 0.714887227614721
42 0.710324345032374
43 0.705266364415487
44 0.713340729475021
45 0.710121959447861
46 0.713599592447281
47 0.704361832141876
48 0.703318081299464
49 0.707208555936813
50 0.708930063247681
51 0.708541176716487
52 0.706024523576101
53 0.706982847054799
54 0.713527637720108
55 0.712104618549347
56 0.715284230311712
57 0.714014929533005
58 0.718396516640981
59 0.712120225032171
60 0.70300657749176
61 0.712330993016561
62 0.705399974187215
63 0.716652369499207
64 0.711654380957286
65 0.715257356564204
66 0.709045793612798
67 0.710406134525935
68 0.713286699851354
69 0.713970551888148
70 0.714134921630224
71 0.708393577734629
72 0.71591382821401
73 0.708847548564275
74 0.716458161671956
75 0.717960963646571
76 0.716552470127741
77 0.71113104224205
78 0.718919990460078
79 0.707574218511581
80 0.714642171065013
81 0.714416901270548
82 0.706916064023972
83 0.711730722586314
84 0.718201941251755
85 0.708433689673742
86 0.716953217983246
87 0.710274974505107
88 0.712638219197591
89 0.712405643860499
90 0.717525462309519
91 0.716002144416173
92 0.711754639943441
93 0.713744324445724
94 0.711945599317551
95 0.720099818706512
96 0.712131476402283
97 0.712278308471044
98 0.713869651158651
99 0.712531715631485
};
\addplot [semithick, forestgreen4416044]
table {%
0 1.55465695460637
1 3.54276961485545
2 6.84717055956523
3 11.9553510348002
4 19.6561587651571
5 30.3932488759359
6 45.8300317128499
7 65.1476365407308
8 89.9160723368327
9 120.983692932129
10 158.812667338053
11 201.779645284017
12 252.178554789225
13 311.822377522786
14 377.131458536784
15 452.17714436849
16 537.134820556641
17 626.936676025391
18 724.715093994141
19 833.412182617187
20 933.673201497396
21 1062.58614298503
22 1190.99056803385
23 1323.37373046875
24 1475.06894124349
25 1623.04679361979
26 1796.58349609375
27 1975.16912434896
28 2162.06663411458
29 2351.95164794922
30 2534.69059651693
31 2714.21853841146
32 2902.59532877604
33 3122.97405598958
34 3328.1301188151
35 3555.73961588542
36 3765.49230957031
37 3952.48976236979
38 4180.17137858073
39 4398.964453125
40 4583.77263997396
41 4859.77096354167
42 5055.92880045573
43 5308.26098632812
44 5534.4049235026
45 5709.12517903646
46 5985.44622395833
47 6210.53883463542
48 6397.46751302083
49 6480.77866210937
50 6707.73958333333
51 6937.21790364583
52 7081.87277018229
53 7304.57184244792
54 7511.70415039062
55 7744.83385416667
56 7863.42859700521
57 8010.54152018229
58 8161.38434244792
59 8352.13855794271
60 8576.60143229167
61 8720.91819661458
62 8963.25335286458
63 9117.78852539062
64 9383.36124674479
65 9600.45442708333
66 9750.65011393229
67 9912.16370442708
68 10039.796891276
69 10187.0822428385
70 10414.3989257812
71 10564.920296224
72 10784.1141113281
73 10934.0772298177
74 11117.7188639323
75 11350.2178710937
76 11500.594905599
77 11680.1168945312
78 11773.5132161458
79 11925.4419596354
80 12148.5900065104
81 12169.8748046875
82 12290.8735026042
83 12554.8247395833
84 12507.3008463542
85 12676.0856770833
86 12885.6792317708
87 13032.0627278646
88 13163.2922851563
89 13345.2062825521
90 13582.3030924479
91 13619.6559244792
92 13966.8358723958
93 14086.3044270833
94 14076.3888671875
95 14291.4867838542
96 14588.037890625
97 14730.4602864583
98 15068.0731119792
99 15193.1791666667
};
\addplot [semithick, crimson2143940]
table {%
0 1.54190982580185
1 3.52118768692017
2 5.78630299568176
3 7.67343028386434
4 9.32714433670044
5 10.8580426216125
6 12.2028846740723
7 13.5414336204529
8 14.8556263923645
9 16.017956829071
10 17.1215937614441
11 18.1431117375692
12 19.2817598342896
13 20.121834564209
14 21.0518109639486
15 21.8852607727051
16 22.524635887146
17 23.5350463867188
18 24.1527964274089
19 24.9533289591471
20 25.9407367070516
21 26.3416049957275
22 27.1657480875651
23 27.5096417109172
24 28.3973735173543
25 28.8585123697917
26 29.5232582092285
27 29.9796120325724
28 30.5138794581095
29 31.1672804514567
30 31.6722736358643
31 32.1286259333293
32 32.4027076085409
33 32.9874686559041
34 33.7719740549723
35 34.2146395365397
36 34.6404315948486
37 35.3181519826253
38 35.8040141423543
39 36.2991956710815
40 36.5904066085815
41 37.1327234903971
42 37.5759311040242
43 38.1746032714844
44 38.2666483561198
45 38.9697507222494
46 39.2302190144857
47 39.9760008494059
48 40.1718897501628
49 40.8396077473958
50 41.1574705759684
51 41.4496417999268
52 42.11405321757
53 42.1532773335775
54 42.8213433583577
55 43.2922027587891
56 43.3367314656576
57 43.3879510243734
58 43.8423197428385
59 44.7495100657145
60 44.7827958424886
61 45.1918746948242
62 46.013836924235
63 46.3216091156006
64 46.8736853281657
65 47.1226577758789
66 47.4211101531982
67 47.7491437276204
68 47.9290562947591
69 48.2311442057292
70 48.5239919026693
71 48.7080121358236
72 49.0425851186117
73 49.4799237569173
74 49.343957010905
75 49.7897899627686
76 50.324275080363
77 50.5796007792155
78 51.2719327290853
79 51.2950219472249
80 51.5075139363607
81 52.4433975219727
82 52.6479281107585
83 52.910989634196
84 53.265039952596
85 53.6047564188639
86 53.7609555562337
87 54.2889787038167
88 54.3406288146973
89 54.391259765625
90 54.7481624603272
91 54.9078276316325
92 55.0772015889486
93 55.4503991444906
94 56.1480810801188
95 56.5388488769531
96 56.1639583587646
97 56.6805483500163
98 56.5415815989176
99 56.8630076090495
};

\end{axis}

\end{tikzpicture}

%% file: Figures/Plot3loss.tex
\begin{tikzpicture}

\definecolor{crimson2143940}{RGB}{214,39,40}
\definecolor{darkgray176}{RGB}{176,176,176}
\definecolor{darkorange25512714}{RGB}{255,127,14}
\definecolor{forestgreen4416044}{RGB}{44,160,44}
\definecolor{lightgray204}{RGB}{204,204,204}
\definecolor{steelblue31119180}{RGB}{31,119,180}

\begin{axis}[
ymode = log,
yticklabels={,.01,.1},
name = plot1,
    width=3cm,  
    height=3cm,  
    scale only axis, 
    xlabel={Epochs},
legend cell align={center},
legend style={
  fill opacity=0.8,
  draw opacity=1,
  text opacity=1,
  at={(1.25,1.5)},
  anchor=north,
  draw=lightgray204,
  legend columns = -1
  mark=*
},
legend image code/.code={
        \draw[mark options={scale=1}, mark=*] plot coordinates {(0.3cm,0cm)}; 
    },
tick align=outside,
tick pos=left,
x grid style={darkgray176},
xmajorgrids,
xmin=0, xmax=100,
xtick style={color=black},
y grid style={darkgray176},
ylabel={},
title={MSE-Loss (\ref{equ:MSELoss})},
ymajorgrids,
ymin=0, 
ymax=0.7,
ytick style={color=black},
]
\addplot [semithick, steelblue31119180]
table {%
0 0.548415012225509
1 0.260596112176776
2 0.199333541236818
3 0.170763516455889
4 0.132872804656625
5 0.0918383099529892
6 0.0567078236807138
7 0.0406543923430145
8 0.0363865633243695
9 0.00801322893449105
10 0.00752538197487593
11 0.00752483715652488
12 0.00752032087184489
13 0.00752312026103027
14 0.00752570083457977
15 0.0075128314960748
16 0.00752777750534005
17 0.00752507122093812
18 0.00752093595499173
19 0.00752727594855241
20 0.00752138263569213
21 0.00751542641315609
22 0.00752721480140462
23 0.007520337237278
24 0.00752563506388105
25 0.00753043382498436
26 0.00752470010262914
27 0.00752017146185972
28 0.00752917782124132
29 0.00752881077281199
30 0.00751396194892004
31 0.00751694584148936
32 0.00752815241133794
33 0.00752066366304643
34 0.00752370889740996
35 0.0075245041269809
36 0.00752782307215966
37 0.00752837221557274
38 0.00752370861545205
39 0.00753300691070035
40 0.00752386529557407
41 0.00752627824433148
42 0.00752069049724378
43 0.00751482116989791
44 0.00750940577406436
45 0.00752902232459746
46 0.00751835922966711
47 0.00752379640331492
48 0.00751147268363275
49 0.00752187384222634
50 0.00753117379080504
51 0.00753021406545304
52 0.00752012984850444
53 0.00753299265610985
54 0.00752129695983604
55 0.00751948116952553
56 0.00752447230764665
57 0.00751921636611223
58 0.00751629655295983
59 0.00751851693820208
60 0.00751744952122681
61 0.00751953963609412
62 0.00752549178292975
63 0.00751764936535619
64 0.00752256043767557
65 0.00752234915737063
66 0.00752038859669119
67 0.00753002483327873
68 0.00752449164609425
69 0.00752483649901114
70 0.00751591556938365
71 0.00752251353464089
72 0.00751609408552758
73 0.00751590899424627
74 0.00751730271382257
75 0.00751301128650084
76 0.00752065231674351
77 0.0075163100764621
78 0.00752277068328112
79 0.0075161181895528
80 0.00752222670963965
81 0.00751390267442912
82 0.00751506595872343
83 0.0075175651377067
84 0.00751211134647019
85 0.00751949909585528
86 0.00753027279488742
87 0.00751712884264998
88 0.00751450412650593
89 0.00751784336776473
90 0.0075144024589099
91 0.00751551885413937
92 0.0075192875280045
93 0.0075173814680893
94 0.00750582735124044
95 0.00751378774479963
96 0.00751508517912589
97 0.00751278982986696
98 0.00750667718774639
99 0.00751805932610296
};
\addlegendentry{$SQ_\text{mSTE}$}
\addplot [semithick, darkorange25512714]
table {%
0 0.55088772341609
1 0.364599603518844
2 0.368866781711578
3 0.377288254261017
4 0.388566744953394
5 0.401835393205285
6 0.414005573749542
7 0.424335187733173
8 0.433307941555977
9 0.440919484496117
10 0.447635944843292
11 0.453682038679719
12 0.458860293194652
13 0.463521628290415
14 0.467869847580791
15 0.471703230798244
16 0.475401790723205
17 0.478585365355015
18 0.481512870833278
19 0.484425481408834
20 0.487144548967481
21 0.489412767782807
22 0.491777189701796
23 0.493936263769865
24 0.495904034361243
25 0.497762332350016
26 0.499550337582827
27 0.501130907699466
28 0.502885586768389
29 0.504318796187639
30 0.505741798982024
31 0.507270608186722
32 0.508436697110534
33 0.509720933899283
34 0.510865337014198
35 0.511979941368103
36 0.513019118696451
37 0.513959438443184
38 0.514955953866243
39 0.516012669444084
40 0.516867266148329
41 0.517744843870401
42 0.51857935076952
43 0.519320396065712
44 0.519961907505989
45 0.520658597141504
46 0.521395146280527
47 0.521903126507998
48 0.522700056582689
49 0.52323027163744
50 0.523826842874289
51 0.524266020983458
52 0.524801029682159
53 0.525233933061361
54 0.525834440082312
55 0.526259167104959
56 0.526617806315422
57 0.527145440518856
58 0.527537268251181
59 0.527885058254004
60 0.528327745348215
61 0.52878401529789
62 0.529222670078278
63 0.529591565400362
64 0.529899104863405
65 0.530336342602968
66 0.530688235938549
67 0.530990009218454
68 0.531487732738256
69 0.531842892467976
70 0.532275775909424
71 0.532542983800173
72 0.532737836956978
73 0.533120610415936
74 0.533493458986282
75 0.533814318478107
76 0.534058399617672
77 0.534519419878721
78 0.534666323184967
79 0.534989660829306
80 0.535261952996254
81 0.535537146002054
82 0.535684410214424
83 0.535837888389826
84 0.536177230507135
85 0.53638651779294
86 0.536517620921135
87 0.536831999242306
88 0.536970202863216
89 0.537155061453581
90 0.537432105869055
91 0.53773717880249
92 0.53779073548317
93 0.538084347188473
94 0.538191051691771
95 0.53855269575119
96 0.53880144482851
97 0.538929655462503
98 0.5389938133955
99 0.539319223552942
};
\addlegendentry{$SQ_\text{STE}$}
\addplot [semithick, forestgreen4416044]
table {%
0 0.549000259801745
1 0.299871038094163
2 0.242189251571894
3 0.207308624267578
4 0.197074625216424
5 0.185634178325534
6 0.182380964674056
7 0.178326630882919
8 0.175201226659119
9 0.173283803023398
10 0.172094418533146
11 0.171343371406198
12 0.170765378408134
13 0.169521951183677
14 0.167103111922741
15 0.157875367105007
16 0.154614733763039
17 0.153081586278975
18 0.150951494216919
19 0.149476350657642
20 0.147199265860021
21 0.144910294152796
22 0.143213532932103
23 0.14128946762532
24 0.137701354525983
25 0.136567906014621
26 0.134132204532623
27 0.132960757054389
28 0.132355139888823
29 0.131965770825744
30 0.131745044507086
31 0.131587471209466
32 0.13147689602524
33 0.131349040202796
34 0.131103035457432
35 0.130517799466848
36 0.130464115925133
37 0.130513427942991
38 0.130506151787937
39 0.130541185207665
40 0.130534353867173
41 0.130492865376174
42 0.130522380650043
43 0.130526529125869
44 0.130550112396479
45 0.130531256340444
46 0.130503811851144
47 0.130515778191388
48 0.130510073944926
49 0.130536538757384
50 0.130526625037193
51 0.130501929484308
52 0.130520934864879
53 0.130509537316859
54 0.130512963078916
55 0.130522072032094
56 0.130513535268605
57 0.130511100739241
58 0.130539780512452
59 0.130513597756624
60 0.130501266375184
61 0.130528594397008
62 0.130528959475458
63 0.130543941237032
64 0.130528621524572
65 0.130521844916046
66 0.130523137196898
67 0.130502953402698
68 0.130502688154578
69 0.130511386707425
70 0.130524727001786
71 0.130521879352629
72 0.130487557888031
73 0.130523864358664
74 0.130489279292524
75 0.130496246568859
76 0.130508339770138
77 0.130523177161813
78 0.130494157969952
79 0.130501418083906
80 0.130529735192657
81 0.130518131420016
82 0.130508684828877
83 0.130500476054847
84 0.130506072625518
85 0.130520362116396
86 0.13052210765332
87 0.130500621564686
88 0.130523995541036
89 0.130495601221919
90 0.130508861780167
91 0.130531919181347
92 0.130510828539729
93 0.130509101711214
94 0.130512753784657
95 0.130492839120328
96 0.130500469848514
97 0.130490967720747
98 0.130495262987912
99 0.130493671834469
};
\addlegendentry{$SQ_\text{STE}^\text{CL}$}
\addplot [semithick, crimson2143940]
table {%
0 0.544153192952275
1 0.248425985530019
2 0.164691873550415
3 0.128648992173374
4 0.0838899204917252
5 0.0417874237690121
6 0.0385444990582764
7 0.0385049095787108
8 0.0384967143498361
9 0.0385195656865835
10 0.0385160425622016
11 0.0385357322134078
12 0.038534109480679
13 0.0385317649245262
14 0.0385209232810885
15 0.0385310827456415
16 0.0384953307788819
17 0.0384952042140067
18 0.0385053486600518
19 0.0385036334935576
20 0.0384773015771061
21 0.0286449612174183
22 0.00934916036436334
23 0.00934722829656675
24 0.00934804730443284
25 0.00934688501665369
26 0.00934892255626619
27 0.00935017336159944
28 0.00935114128375426
29 0.00934771656431258
30 0.00934540102351457
31 0.00935453312192112
32 0.00935013657994568
33 0.00934792333096266
34 0.00934989839419723
35 0.00935237167496234
36 0.00934469219530001
37 0.0093483248539269
38 0.00935291248885915
39 0.00934915019385517
40 0.00934571000235155
41 0.00934837823081762
42 0.00934899662155658
43 0.00935033038724214
44 0.00934421865548938
45 0.00934903470706195
46 0.00934945413237438
47 0.00934787023393437
48 0.00934440769907087
49 0.00934540874511004
50 0.00935275487368926
51 0.00935078661795706
52 0.00934717530291528
53 0.009345349971205
54 0.00935187224624679
55 0.00934593824390322
56 0.00934900092938915
57 0.00934595982916653
58 0.00934418482193723
59 0.00935198008688167
60 0.00935088515095413
61 0.00934546708734706
62 0.00934788803616539
63 0.00934711833763868
64 0.00935056802397594
65 0.00934861720679328
66 0.00934991103224456
67 0.00934911843948066
68 0.00934640683792532
69 0.00934659997047856
70 0.00935074001271278
71 0.00934409358212724
72 0.00934975521126762
73 0.00935268123727292
74 0.00934827431710437
75 0.00934959468618035
76 0.0093522864584811
77 0.00935016064764932
78 0.00935106620984152
79 0.00934808949381113
80 0.00934725726908073
81 0.00935152086615562
82 0.00934670987585559
83 0.00935189650719985
84 0.00934574100933969
85 0.00934906351426616
86 0.00935401551006362
87 0.00935002066288143
88 0.00934754252713174
89 0.00934870791714638
90 0.00935289056319743
91 0.00934692508447915
92 0.00934829985210672
93 0.00935069577442482
94 0.00934891671128571
95 0.00934632327081636
96 0.00935236550169066
97 0.00934587485063821
98 0.00935068418178707
99 0.00934836278343573
};
\addlegendentry{$SQ_\text{mSTE}^\text{CL}$}
\end{axis}

\begin{axis}[
        at={(plot1.east)}, 
    anchor=west,
    width=3cm,  
    height=3cm,  
     xshift=1.1cm, 
    scale only axis, 
  title={MA-E (\ref{equ:MAE})},
log basis y={10},
tick align=outside,
tick pos=left,
x grid style={darkgray176},
xlabel={Epochs},
xmajorgrids,
xmin=0, xmax=100,
xtick style={color=black},
y grid style={darkgray176},
ymajorgrids,
ymin=0.37772735089694, ymax=302220.574977491,
ymode=log,
ytick style={color=black}
]
\addplot [semithick, steelblue31119180]
table {%
0 0.851375661293666
1 0.806172464291255
2 0.812832925717036
3 0.827050844828288
4 0.828881853818893
5 0.839782045284907
6 0.837877966960271
7 0.839749137560526
8 0.841935312747955
9 0.827198886871338
10 0.829452168941498
11 0.828425870339076
12 0.828687649965286
13 0.828649872541428
14 0.828402545054754
15 0.829283454020818
16 0.829582351446152
17 0.827392035722733
18 0.827296247084936
19 0.826385231812795
20 0.825873935222626
21 0.826272092262904
22 0.829095045725505
23 0.828043138980865
24 0.829359706242879
25 0.825448644161224
26 0.831103394428889
27 0.82834423383077
28 0.827774888277054
29 0.826178681850433
30 0.82883224884669
31 0.832357136408488
32 0.828405461708705
33 0.827094235022863
34 0.827335288127263
35 0.828002520402273
36 0.827962724367778
37 0.8313736140728
38 0.827159323294957
39 0.828334468603134
40 0.828757450977961
41 0.829585385322571
42 0.828349604209264
43 0.8305335521698
44 0.826499412457148
45 0.825795143842697
46 0.826147758960724
47 0.826950929562251
48 0.825480167071025
49 0.829947408040365
50 0.823849985996882
51 0.825413328409195
52 0.828409097592036
53 0.825286384423574
54 0.823809057474136
55 0.83239281574885
56 0.826397462685903
57 0.826078220208486
58 0.82554986278216
59 0.829379457235336
60 0.828008983532588
61 0.823278741041819
62 0.823613027731578
63 0.828036558628082
64 0.826531108220418
65 0.822369156281153
66 0.823448852698008
67 0.829317446549733
68 0.828767079114914
69 0.822965995470683
70 0.824071802695592
71 0.826074622074763
72 0.823374515771866
73 0.828081520398458
74 0.827080847819646
75 0.826028686761856
76 0.827883988618851
77 0.825355138381322
78 0.821851378679275
79 0.822457506259282
80 0.829773046573003
81 0.826981800794601
82 0.827518127361933
83 0.82455484867096
84 0.826452980438868
85 0.828788681825002
86 0.825748177369436
87 0.827113646268845
88 0.823399613300959
89 0.825183908144633
90 0.82611332933108
91 0.826250165700912
92 0.827026361227036
93 0.826306056976318
94 0.828172109524409
95 0.826435246070226
96 0.824748667081197
97 0.824827434619268
98 0.827027906974157
99 0.82360468506813
};
\addplot [semithick, darkorange25512714]
table {%
0 1.52191224098206
1 4.62942314147949
2 10.7700797080994
3 20.8528925577799
4 36.3747828801473
5 58.6029336293538
6 88.5621025085449
7 125.331676737467
8 172.717534383138
9 231.091755167643
10 297.690074666341
11 374.748385620117
12 470.715333048503
13 584.981895955404
14 705.257332356771
15 840.939318847656
16 996.562082926432
17 1164.62107543945
18 1361.14012451172
19 1563.32934977214
20 1795.80465087891
21 2030.06355794271
22 2333.39939778646
23 2611.14085286458
24 2954.9336710612
25 3288.2523844401
26 3671.10323079427
27 4043.10830891927
28 4452.31261393229
29 4877.00172526042
30 5379.12698567708
31 5833.47943522135
32 6415.72179361979
33 7065.05356445312
34 7571.12750651042
35 8357.25302734375
36 8967.70460611979
37 9765.99383138021
38 10491.2435546875
39 11156.7108072917
40 12068.3065917969
41 12973.3904459635
42 13800.2222981771
43 14856.6091145833
44 15786.3408854167
45 16714.0112630208
46 17757.7693684896
47 18953.1794596354
48 20258.6350585938
49 21322.02734375
50 22663.4591471354
51 23886.9681640625
52 25237.6015299479
53 26856.3376953125
54 28415.2879557292
55 29912.9180338542
56 31284.2891276042
57 32826.7267578125
58 34398.5539713542
59 36384.1580729167
60 38069.0947265625
61 39918.037890625
62 42019.1438151042
63 44091.3958984375
64 45745.9811848958
65 48158.7407552083
66 50503.2204427083
67 52731.0869791667
68 54221.2240885417
69 57230.4666666667
70 60084.062109375
71 62508.45625
72 65623.8298177083
73 66958.82265625
74 69791.838671875
75 72859.9966145833
76 76534.0283854167
77 78393.4149739583
78 80455.669921875
79 84537.4856770833
80 88595.2701822917
81 90587.526953125
82 93212.7028645833
83 98404.0579427083
84 100310.597395833
85 104992.432682292
86 107600.389583333
87 111554.88125
88 115896.441666667
89 120097.72578125
90 123141.6984375
91 127448.078385417
92 129743.982552083
93 134410.966927083
94 140104.898958333
95 143944.02421875
96 148692.367708333
97 153510.48828125
98 157471.023958333
99 162929.313541667
};
\addplot [semithick, forestgreen4416044]
table {%
0 1.15177705486616
1 1.49898407856623
2 1.68984109958013
3 1.74100794792175
4 1.81043399969737
5 1.80887944698334
6 1.81068992614746
7 1.80273396571477
8 1.79722727139791
9 1.80120892524719
10 1.79876539707184
11 1.8107432325681
12 1.81065675814947
13 1.82434448401133
14 1.8075327595075
15 1.9066148519516
16 1.90987679560979
17 1.90051119724909
18 1.88925451437632
19 1.88267554442088
20 1.87240897019704
21 1.85495265722275
22 1.84392866690954
23 1.8233410914739
24 1.80795597235362
25 1.81095338265101
26 1.7882888674736
27 1.79436893860499
28 1.7830516854922
29 1.78319657246272
30 1.78521246512731
31 1.78399416208267
32 1.77897806962331
33 1.78484630187352
34 1.78965667883555
35 1.78724503119787
36 1.77722928126653
37 1.78671556711197
38 1.78292543490728
39 1.77811615069707
40 1.77549083630244
41 1.78069831530253
42 1.77355553309123
43 1.77956559658051
44 1.78367565870285
45 1.77791599432627
46 1.77011037667592
47 1.7834709127744
48 1.77421322266261
49 1.77529484033585
50 1.78160897095998
51 1.77904099623362
52 1.77556119362513
53 1.78328501383464
54 1.78391941785812
55 1.78135044972102
56 1.7737685362498
57 1.78358612060547
58 1.77882724603017
59 1.7719974120458
60 1.77522218624751
61 1.77480307420095
62 1.77991149822871
63 1.77920944690704
64 1.77348047494888
65 1.78633494377136
66 1.77663913567861
67 1.77468335231145
68 1.78009341160456
69 1.78007904291153
70 1.77263360420863
71 1.77128464778264
72 1.7840326944987
73 1.77595264911652
74 1.77936276594798
75 1.78149652481079
76 1.78397656679153
77 1.77532234191895
78 1.78522520860036
79 1.77818075418472
80 1.78747274080912
81 1.77702915668488
82 1.77857061624527
83 1.77995453675588
84 1.7965615272522
85 1.77666591008504
86 1.78238356908162
87 1.77309301694234
88 1.78007437785467
89 1.77789382537206
90 1.78000446160634
91 1.78402543465296
92 1.78007783095042
93 1.77365928490957
94 1.78022813002268
95 1.77402989466985
96 1.78552823464076
97 1.77882620096207
98 1.78523919582367
99 1.7766180674235
};
\addplot [semithick, crimson2143940]
table {%
0 0.80241298476855
1 0.779247009754181
2 0.780947726964951
3 0.799171262979507
4 0.811832990248998
5 0.818164143959681
6 0.814376646280289
7 0.817777268091838
8 0.814427242676417
9 0.81681828101476
10 0.818162186940511
11 0.82069628238678
12 0.818000364303589
13 0.819307013352712
14 0.817373565832774
15 0.816734490791957
16 0.819832017024358
17 0.817500774065653
18 0.819126492738724
19 0.818786593278249
20 0.815692134698232
21 0.825246717532476
22 0.821618354320526
23 0.820875112215678
24 0.821256818373998
25 0.820211603244146
26 0.822047559420268
27 0.825259459018707
28 0.82359991868337
29 0.825164367755254
30 0.81974901954333
31 0.825739628076553
32 0.824142911036809
33 0.823180478811264
34 0.823896817366282
35 0.822326036294301
36 0.82332470814387
37 0.820419631401698
38 0.821594359477361
39 0.822244900465012
40 0.821531329552333
41 0.823203843832016
42 0.820670145750046
43 0.823018465439479
44 0.820708586772283
45 0.823641169071198
46 0.823253548145294
47 0.820156538486481
48 0.821354840199153
49 0.822765853007634
50 0.823422499497731
51 0.821060663461685
52 0.821967407067617
53 0.824151565631231
54 0.825473699967066
55 0.821656952301661
56 0.824345552921295
57 0.825614617268244
58 0.824551910161972
59 0.821677666902542
60 0.82259711821874
61 0.821704626083374
62 0.820801309744517
63 0.82317022283872
64 0.824893373250961
65 0.822421731551488
66 0.823324050505956
67 0.823649946848551
68 0.820150379339854
69 0.821622826655706
70 0.821760533253352
71 0.824078822135925
72 0.824826695521673
73 0.823277656237284
74 0.823374440272649
75 0.822636226812998
76 0.821602088212967
77 0.8195450146993
78 0.8201846520106
79 0.822569251060486
80 0.823647185166677
81 0.822407760222753
82 0.820196986198425
83 0.822377115488052
84 0.824412471055984
85 0.823043855031331
86 0.823053475220998
87 0.821783860524495
88 0.821896227200826
89 0.82070717215538
90 0.823118283351262
91 0.821484877665838
92 0.822438844045003
93 0.822589373588562
94 0.820179446538289
95 0.824486400683721
96 0.820729015270869
97 0.823473676045736
98 0.821667681137721
99 0.822946792840958
};
\end{axis}

\end{tikzpicture}

%% file: Figures/PlotXcodecf_norms.tex
\begin{tikzpicture}

\definecolor{darkgray176}{RGB}{176,176,176}
\definecolor{darkorange25512714}{RGB}{255,127,14}
\definecolor{forestgreen4416044}{RGB}{44,160,44}
\definecolor{lightgray204}{RGB}{204,204,204}
\definecolor{steelblue31119180}{RGB}{31,119,180}

\begin{axis}[
    anchor=west,
    width=3cm,  
    height=3cm, 
    legend cell align={center},
legend style={
  fill opacity=0.8,
  draw opacity=1,
  text opacity=1,
  at={(0.5,1.5)},
  anchor=north,
  draw=lightgray204,
  legend columns = -1
  mark=*
},
    legend image code/.code={
        \draw[mark options={scale=1}, mark=*] plot coordinates {(0.3cm,0cm)}; 
    },
     xshift=1.5cm, 
    scale only axis, 
 title={MA-E (\ref{equ:MAE})},
log basis y={10},
tick align=outside,
tick pos=left,
x grid style={darkgray176},
xlabel={Epochs},
xmajorgrids,
xmin=0, xmax=30,
xtick style={color=black},
y grid style={darkgray176},
ymajorgrids,
ymin=0.001, ymax=10 0000000000000000,
ymode=log,
ytick style={color=black}
]
\addplot [semithick, steelblue31119180]
table {%
0 220920456.586761
1 1336126358.49353
2 63976024621.5685
3 148516052586.49
4 320987298775.846
5 517504113655.364
6 1051341255736.73
7 1081520208140.33
8 1093097830481.14
9 1277915579649.94
10 1303095408281.98
11 3659127229661.41
12 10826622393003.5
13 11293220986621
14 11914486941395.9
15 15450026732433.4
16 36711791912633.6
17 102606070269964
18 115514332066640
19 197521567035481
20 206259469442689
21 263319425419315
22 338359262263218
23 791294460225600
24 910537334295859
25 1.37393198631451e+15
26 4.65460455963129e+15
27 5.52125460382298e+15
28 1.48697738640337e+16
29 1.58119710192262e+16
30 1.69636369374388e+16
31 2.13666669903096e+16
};
\addlegendentry{$\text{NA}_{\text{det}}$}
\addplot [semithick, darkorange25512714, dashed]
table {%
0 0.0583785891631658
1 0.0583785891631658
2 0.0583785891631658
3 0.0583785891631658
4 0.0583785891631658
5 0.0583785891631658
6 0.0583785891631658
7 0.0583785891631658
8 0.0583785891631658
9 0.0583785891631658
10 0.0583785891631658
11 0.0583785891631658
12 0.0583785891631658
13 0.0583785891631658
14 0.0583785891631658
15 0.0583785891631658
16 0.0583785891631658
17 0.0583785891631658
18 0.0583785891631658
19 0.0583785891631658
20 0.0583785891631658
21 0.0583785891631658
22 0.0583785891631658
23 0.0583785891631658
24 0.0583785891631658
25 0.0583785891631658
26 0.0583785891631658
27 0.0583785891631658
28 0.0583785891631658
29 0.0583785891631658
30 0.0583785891631658
31 0.0583785891631658
};
\addlegendentry{NA}
\addplot [semithick, forestgreen4416044, dashed]
table {%
0 0.34256650745833
1 0.34256650745833
2 0.34256650745833
3 0.34256650745833
4 0.34256650745833
5 0.34256650745833
6 0.34256650745833
7 0.34256650745833
8 0.34256650745833
9 0.34256650745833
10 0.34256650745833
11 0.34256650745833
12 0.34256650745833
13 0.34256650745833
14 0.34256650745833
15 0.34256650745833
16 0.34256650745833
17 0.34256650745833
18 0.34256650745833
19 0.34256650745833
20 0.34256650745833
21 0.34256650745833
22 0.34256650745833
23 0.34256650745833
24 0.34256650745833
25 0.34256650745833
26 0.34256650745833
27 0.34256650745833
28 0.34256650745833
29 0.34256650745833
30 0.34256650745833
31 0.34256650745833
};
\addlegendentry{no Quantizer}
\end{axis}

\end{tikzpicture}